\documentclass[a4paper,fleqn,usenatbib]{mnras}

\usepackage{graphicx}
\usepackage{amsmath}
\usepackage{amssymb}
\usepackage{newtxtext,newtxmath}
\usepackage{Times}

\usepackage[T1]{fontenc}
\usepackage{ae,aecompl}

\usepackage[dvipsnames]{xcolor}
\usepackage{lipsum}
\usepackage{natbib}
\usepackage{multirow}
\usepackage{booktabs}

\usepackage{soul}

\title[UV compactness and morphologies of LAEs from $z\sim2-6$]{On the UV compactness and morphologies of typical Lyman-$\boldsymbol{\alpha}$ emitters from $\boldsymbol{z\sim2}$ to $\boldsymbol{z\sim6}$}

\author[A. Paulino-Afonso et al.]{Ana Paulino-Afonso$^{1,2,3}$\thanks{E-mail: aafonso@oal.ul.pt},
David Sobral$^{3,4}$,
Bruno Ribeiro$^{5}$,
Jorryt Matthee$^{4}$,
\newauthor
S\'ergio Santos$^{3}$,
Jo\~ao Calhau$^{3}$,
Alex Forshaw$^{3}$,
Andrea Johnson$^{3}$,
Joanna Merrick$^{3}$,
\newauthor
Sara P\'erez$^{1,2,3}$,
Oliver Sheldon$^{3}$
\\
$^{1}$Instituto de Astrof\'isica e Ci\^encias do Espa\c{c}o, Universidade de Lisboa, OAL, Tapada da Ajuda, PT1349-018 Lisboa, Portugal\\
$^{2}$Departamento de F\'isica, Faculdade de Ci\^encias, Universidade de Lisboa, Edif\'icio C8, Campo Grande, PT1749-016 Lisboa, Portugal\\
$^{3}$Department of Physics, Lancaster University, Lancaster, LA1 4YB, UK\\
$^{4}$Leiden Observatory, Leiden University, P.O. Box 9513, NL-2300 RA Leiden, The Netherlands\\
$^{5}$Centro de Computa\c{c}\~{a}o Gr\'afica, CVIG,  Campus de Azur\'em, PT4800-058 Guimar\~{a}es, Portugal}

\date{Accepted 2018 January 30. Received 2017 November 6; in original form 2017 July 25}

\pubyear{2018}

\begin{document}

\label{firstpage}
\pagerange{\pageref{firstpage}--\pageref{lastpage}}
\maketitle

\begin{abstract}
We investigate the rest-frame UV morphologies of a large sample of Lyman-$\alpha$ emitters (LAEs) from $z\sim2$ to $z\sim6$, selected in a uniform way with 16 different narrow- and medium-bands over the full COSMOS field.
We use 3045 LAEs with HST coverage in a stacking analysis and find that they have $M_\mathrm{UV}\sim-20$, below $M_\mathrm{UV}^\ast$ at these redshifts. We also focus our analysis on a subsample of 780 individual galaxies with $i_\mathrm{AB}<25$ for which GALFIT converges for 429 of them.
The individual median size ($r_e\sim1$ kpc), ellipticities (slightly elongated with $(b/a)\sim0.45$), S\'ersic index (disk-like with $n\lesssim2$) and light concentration (comparable to that of disk or irregular galaxies, with $C\sim2.7$) of LAEs show mild evolution from $z\sim2$ to $z\sim6$. LAEs with the highest rest-frame equivalent widths (EW) are the smallest/most compact ($r_e\sim0.8$ kpc, compared to $r_e\sim1.5$ kpc for the lower EW LAEs). When stacking our samples in bins of fixed Ly$\alpha$ luminosity and Ly$\alpha$ EW we find evidence for redshift evolution in $n$ and $C$, but not in galaxy sizes. The evolution seems to be stronger for LAEs with $25 <EW<100$ \AA. When compared to other SFGs, LAEs are found to be smaller at all redshifts. The difference between the two populations changes with redshift, from a factor of $\sim1$ at $z\gtrsim5$ to SFGs being a factor of $\sim2-4$ larger than LAEs for $z\lesssim2$. This means that at the highest redshifts, where typical sizes approach those of LAEs, the fraction of galaxies showing Ly$\alpha$ in emission (and with a high Ly$\alpha$ escape fraction) should be much higher, consistent with observations. 
\vspace{2em}
\end{abstract}



\bibpunct[ ]{(}{)}{;}{a}{}{,}


\section{Introduction}\label{section:introduction}

In the $\Lambda$-Cold Dark Matter framework, galaxies form through the coalescence of small clumps of material \citep[see e.g.][and references therein]{somerville2015}. This means that the first objects which can be called galaxies are to be young, small and with low stellar mass content. The search for these building blocks of current day galaxies has been pursued intensively in the past decades \citep[see e.g.][]{bromm2011,stark2016}. 

Because of its intrinsic brightness, this search usually explores the presence of the Lyman-$\alpha$ (Ly$\alpha$) emission line \citep[e.g.][]{partridge1967,schaerer2003}. This line can be observed in the optical and near-infrared when emitted from $2<z\lesssim8$ sources and it is proven to be a successful probe to identify and confirm high-redshift galaxies. From narrow-band surveys \citep[e.g.][]{rhoads2000,ouchi2008,ouchi2010,matthee2016,matthee2017a,sobral2017} to spectroscopic detection and confirmation of high-redshift candidates \citep[e.g.][]{martin2004,cassata2011,ono2012,bacon2015,lefevre2015,wisotzki2016}, we have now access to large samples of young galaxies in the early Universe.

The physical properties of Ly$\alpha$ emitting galaxies (LAEs) have been intensively studied \citep[e.g.][]{erb2006,gawiser2006,gawiser2007,pentericci2007,ouchi2008,lai2008,reddy2008,finkelstein2009,kornei2010,guaita2011,nilsson2011,acquaviva2012,kusakabe2015,oteo2015,hathi2016,matthee2016}. Some works find them to be typically young, with low stellar masses and scarce dust presence \citep[e.g.][]{erb2006,gawiser2006,gawiser2007,pentericci2007,oteo2015}, while others indicate a more diverse population \citep[e.g.][]{shapley2003,lai2008,reddy2008,finkelstein2009,kornei2010,nilsson2011,acquaviva2012,kusakabe2015,hathi2016}. The different properties of this population may be due to their particular selection method, as LAEs are similar to other line-emission selected galaxies at $z\sim2$ and different from colour-selected galaxies at the same redshifts \citep[e.g.][]{oteo2015,hagen2016}. Evolution can also play a role in the different observed properties of LAEs with more evolved galaxies having Ly$\alpha$ emission driven by different mechanisms than those that dominate LAEs at higher redshifts. For e.g., Sobral et al. (in prep.) show a strong increase in the fraction of Active Galactic Nuclei (AGN) in luminous LAEs from $z\sim4-5$ to $z\sim2-3$.

A possible explanation to the diverse nature of LAEs is linked to the complicated nature of the radiative transfer process itself. To escape the region it originated from, Ly$\alpha$ photons are frequently scattered (with random walks up to several kpc) before they escape towards our line of sight \citep[e.g.][]{zheng2011,dijkstra2012,lake2015,gronke2015,gronke2016}. This recurrent scattering increases the chance of the photon to be destroyed through dust absorption \citep[e.g.][]{neufeld1991,laursen2013}. This picture also means that the particular orientation of the emission path relative to the geometrical distribution of gas and dust in the emitting region is important to consider whether or not we are able to observe the line in emission. Some simulations of isolated disk galaxies have shown that the likelihood of observation of Ly$\alpha$ is correlated with the disk inclination relative to our line of sight \citep{verhamme2012,berhens2014}. From an observational perspective, the Ly$\alpha$ escape fraction (ratio of observed to intrinsic flux in emission) is loosely correlated with the galaxies' star formation rate (SFR) and dust attenuation \citep[e.g.][]{hayes2010,hayes2011,atek2014,matthee2016,trainor2016,oyarzun2017}. The column density of H{\sc i} seems to be another physical quantity that determines the rate of escape of Ly$\alpha$ photons \citep[e.g.][]{shibuya2014a,shibuya2014b,henry2015}. It also correlates with equivalent width \citep[EW,][]{sobral2017,verhamme2017} and outflow velocity \citep[e.g.][]{hashimoto2015}.

The complex process of Ly$\alpha$ escape naturally means that obtaining a complete census of the galaxy population at a given epoch is challenging. To understand the mechanisms that allow Ly$\alpha$ photons to escape it may be important to correlate the morphology of star-forming regions traced by UV continuum emission of young stars with the observed from Ly$\alpha$ photons. This will allow one to constrain the geometry requirements for Ly$\alpha$ to escape from galaxies and further our knowledge of population bias when using selections solely based on this emission line. 
To gain insight on the mechanisms of Ly$\alpha$ escape it is thus crucial that we characterize the morphology of these sources.

Several samples of LAEs have been studied in terms of their rest-frame UV morphologies at $z>2$ \citep[e.g.][]{venemans2005,pirzkal2007,overzier2008,taniguchi2009,bond2009,bond2011,bond2012,gronwall2011,jiang2013,kobayashi2016}. In the local Universe, where rest-frame UV observations are scarce, there is one study based on the Ly$\alpha$ Reference Sample \citep[LARS,][though the sample is H$\alpha$ selected]{ostlin2014}, that characterizes the morphology of these sources \citep{guaita2015}. Observations show that LAEs are typically small, often compact objects (half-light radius around 1 kpc), which undergo no evolution in the first 1 to 3 billion years of the Universe \citep[$z\sim2-6$, e.g][]{venemans2005,malhotra2012}. This scenario is in stark contrast with the stronger evolution in galaxy sizes observed in other populations observed at similar epochs such as Lyman-break galaxies (LBGs) and other star-forming galaxies \citep[e.g.][]{ferguson2004,bouwens2004,vanderwel2014,morishita2014,shibuya2016,paulino-afonso2017}. This can potentially be explained due to the low stellar mass nature of LAEs when compared to other galaxies. However, most studies on SFGs explore the size evolution in stellar mass bins and find stronger size evolution nonetheless \citep[e.g.][]{vanderwel2014}.

One interesting property of LAEs is that the Ly$\alpha$ emission region is often found to be more extended (in a diffuse halo) than the stellar UV continuum emission \citep[e.g.][]{rauch2008,matsuda2012,momose2014,matthee2016,wisotzki2016,sobral2017,xue2017}. The process responsible for such observations is thought to be the scattering of photons by neutral H{\sc i} gas around galaxies at high redshift \citep[e.g.][]{zheng2011}, but could also be due to cooling, satellites and fluorescence \citep[e.g.][]{mas-ribas2017}. Additionally, there are evidences for a correlation between Ly$\alpha$ line luminosity and galaxy UV continuum size \citep[e.g.][]{hagen2014}.

\begin{table*}
\centering
\caption[Narrow and Medium Bands used in this paper.]{The full sample of Ly$\alpha$ emitters selected with the 16 narrow- and medium-bands used in this work. The $<z>$ column shows the average redshift for the LAEs that fall in the filter. The $N_\mathrm{LAE}$ column shows the total number of LAEs detected in the NB/IB images. The $N_\mathrm{HST}$ column shows those who are covered by the HST/ACS F814W imaging survey. The $N_\mathrm{HST, i_\mathrm{AB}<25}$ column shows the number of LAEs with available HST data brighter than $i_\mathrm{AB}<25$. The $N_\mathrm{GALFIT, i_\mathrm{AB}<25}$ column shows the number of bright LAEs for which \textsc{GALFIT} has converged. The $M_\mathrm{F814W}$ [stack] column shows the absolute magnitude in the $i$-band of the median stacks (see Section \ref{section:method_stacks}) which should closely trace $M_\mathrm{UV}$. The $\log_{10}\left(L_\mathrm{Ly\alpha}\right)$ shows the median Ly$\alpha$ luminosity of each sample \citep[values derived by][see also Sobral et al., in prep. and P\'erez et al., in prep.]{matthee2016,santos2016,sobral2017}.}
\begin{tabular}{ccccccccc}
\hline
Band & Instrument &  $<z>$ & $N_\mathrm{LAE}$ & $N_\mathrm{HST}$  & $N_\mathrm{HST, i_\mathrm{AB}<25}$ & $N_\mathrm{GALFIT, i_\mathrm{AB}<25}$ &  $M_\mathrm{F814W}$ [stack] & $\log_{10}\left(L_\mathrm{Ly\alpha}\right)$\\
\hline
NB392 & INT/WFC & 2.23 &   159 &   109 &   50 &    26 & $-18.3\pm0.4$ & $42.6\pm0.3$ \\
NB501 & INT/WFC & 3.12 &    45 &    41 &    17 &     6 & $-19.7\pm0.2$ & $42.9\pm0.2$ \\
\hline
NB711 & Subaru/Suprime-CAM & 4.85 &    78 &    59 &     9 &     4 & $-20.0\pm0.1$ & $42.8\pm0.2$ \\
NB816 & Subaru/Suprime-CAM & 5.72 &   192 &   146 &     4 &     1 & $-19.4\pm0.1$ & $42.8\pm0.2$ \\
\hline
IA427 & Subaru/Suprime-CAM & 2.51 &   741 &   591 &   144 &    83 & $-19.2\pm0.2$ & $42.6\pm0.2$ \\
IA464 & Subaru/Suprime-CAM & 2.82 &   311 &   236 &    96 &    45 & $-19.8\pm0.1$ & $42.9\pm0.2$ \\
IA484 & Subaru/Suprime-CAM & 2.98 &   711 &   563 &   137 &    73 & $-19.5\pm0.1$ & $42.8\pm0.2$ \\
IA505 & Subaru/Suprime-CAM & 3.16 &   483 &   375 &   108 &    65 & $-19.8\pm0.1$ & $42.9\pm0.2$ \\
IA527 & Subaru/Suprime-CAM & 3.34 &   641 &   506 &    98 &    54 & $-19.7\pm0.1$ & $42.9\pm0.2$ \\
IA574 & Subaru/Suprime-CAM & 3.72 &    98 &    77 &    26 &    20 & $-20.2\pm0.1$ & $43.0\pm0.2$ \\
IA624 & Subaru/Suprime-CAM & 4.14 &   142 &   111 &    14 &    10 & $-19.9\pm0.1$ & $43.0\pm0.1$ \\
IA679 & Subaru/Suprime-CAM & 4.59 &    79 &    65 &    23 &    16 & $-20.5\pm0.1$ & $43.3\pm0.1$ \\
IA709 & Subaru/Suprime-CAM & 4.84 &    81 &    59 &    20 &    12 & $-20.7\pm0.1$ & $43.2\pm0.1$ \\
IA738 & Subaru/Suprime-CAM & 5.07 &    79 &    61 &    21 &    10 & $-20.5\pm0.1$ & $43.2\pm0.2$ \\
IA767 & Subaru/Suprime-CAM & 5.31 &    33 &    24 &     9 &     3 & $-20.7\pm0.1$ & $43.4\pm0.2$ \\
IA827 & Subaru/Suprime-CAM & 5.81 &    35 &    22 &     4 &     1 & $-20.0\pm0.2$ & $43.4\pm0.1$ \\
\hline
\hline
Total &  & &  3908 &  3045 &   780 &   429 & & \\
\end{tabular}
\label{tab:summary_bands}
\end{table*}

It is still unclear whether LAEs are a special subset of galaxies, if they rather just trace an early phase of galaxy formation or if they are a consequence of different orientation angles from which Ly$\alpha$ photons peer through. To make progress, we have to look at their morphological properties across cosmic time. In addition to that, it is necessary to compare to other galaxy populations (e.g. LBGs, HAEs, SFGs) for an understanding on how these populations are linked.

In this paper, we analyse in a consistent way, from sample selection to analysis, a large sample of LAEs probing the early phases of galaxy assembly from the end of re-ionization ($z\sim6$) to the peak of the cosmic star-formation history ($z\sim2$). We use data from 16 narrow- and medium-band images in the COSMOS field (Sobral et al., in prep.) to quantify the evolution of galaxy structure (sizes, light profile shapes, elongations and concentrations). With this large data set we can investigate with unprecedented accuracy the  evolution of LAEs sizes and connect that to the evolution (or lack thereof) in other morphological properties and contextualize our results within recent results from the literature on morphology of high redshift galaxies.

The paper is organized as follows. In Section \ref{section:data} we describe the data used for the detection and characterization of LAEs that are the object of study in this work. We present our methodology to study the structural parameters of high redshift galaxies in Section \ref{section:methodology}. The results obtained for the LAEs samples are reported in Section \ref{section:results}. We discuss the implications of our results in the context of early galaxy assembly in Section \ref{section:discussion}. Finally, in Section \ref{section:conclusions} we summarize our conclusions. 

Magnitudes are given in the AB system \citep{oke1983}. All the results assume a $\Lambda$-CDM cosmological model with $H_0$$=$70.0$\ \mathrm{km\ s^{-1}Mpc^{-1}}$,$\ \Omega_m$$=$0.3 and $\Omega_\Lambda$$=$0.7.


\section{The sample of Ly$\alpha$ emitters at $z\sim2-6$}\label{section:data}

The use of narrow-band images to target the Ly$\alpha$ line at specific redshift windows has been widely used in recent years \citep[e.g.][]{rhoads2000,ouchi2008,matsuda2011,konno2014,konno2016,trainor2016,santos2016,matthee2016,sobral2017}. In this paper we use a dataset obtained with the Wide Field Camera at the Isaac Newton Telescope (WFC/INT) and with the Suprime-Cam at Subaru Telescope that cover the full COSMOS field \citep[see][]{scoville2007}.  

We analyse a sample of $\sim$4000 Ly$\alpha$-selected galaxies spanning a wide redshift range of $z\sim2-6$ \citep[SC4K,][and see also Sobral et al., in prep. and P\'erez et al., in prep.]{santos2016,sobral2017}. The sources were detected using a compilation of 16 narrow- and medium-band images taken with the Subaru and the Isaac Newton telescopes. Briefly, sources were classified as Ly$\alpha$ emitters if they satisfied all the following conditions: 1) significant detection in a narrow/medium band with rest-frame equivalent width cuts of 25/50\AA, respectively; 2) presence of a Lyman break blue-ward of the respective narrow/medium band; 3) no strong red colour in the near-infrared, which is typical of a red star or lower redshift interlopers. For the full selection criteria we refer to Sobral et al. (in prep.) and P\'erez et al. (in prep.).

To assess the potential contamination within our sample of LAE candidates we have compiled a set of spectroscopic redshifts from three catalogues - 56 from COSMOS \citep{ilbert2013}, 83 from VUDS \citep[private communication,][]{lefevre2015} and 4 from MOSDEF \citep{kriek2015} - of which 7 objects overlap. We have computed $\delta z = z_\mathrm{NB}-z_\mathrm{spec}$ and find that $\sim$15\% (20 out of 133) of our candidates have $|\delta z|>0.15$ \citep[meaning $\sim$85\% are spectroscopically confirmed LAEs, in agreement with][]{sobral2017}.  We find no significant dependence of the contamination rate on redshift, Ly$\alpha$ luminosity or Ly$\boldsymbol{\alpha}$ equivalent width. 

	\subsection{Sample properties}

We summarize in Appendix \ref{sec:general_properties} the general sample properties plotting the individual band distributions in  Ly$\alpha$ luminosity, $EW_{0}(\mathrm{Ly\alpha})$, and $i_\mathrm{AB}$. We have a total of 3908 LAE candidates over a contiguous area of 2 deg$^2$ in the COSMOS field. These emitters have Ly$\alpha$ luminosities of $10^{42-44}\mathrm{erg\ s^{-1}}$ (see Figure \ref{fig:LAEs_samples_lum}). We note that due to the relatively constant flux limit, the Ly$\alpha$ luminosity limit becomes higher with redshift and we can only detect galaxies above $10^{43}\mathrm{erg\ s^{-1}}$ for $z\gtrsim4$ in the medium bands and $10^{42.5}\mathrm{erg\ s^{-1}}$ for the narrow bands. Despite narrow and intermediate bands having different equivalent width cuts (25 and 50\AA, respectively), we show in Figure \ref{fig:LAEs_samples_ew} that both methods yield similar distributions and the majority of narrow-band selected sources are found at $EW_{0}(\mathrm{Ly\alpha})>50$\AA. To get reliable individual morphological measurements in the rest-frame UV we require a sufficiently bright magnitude in the F814W band and we show in Figure \ref{fig:LAEs_samples_mag} the $i_\mathrm{AB}$ distribution of each sample. We note that we miss a significant fraction of LAEs ($\sim$75\% of those with HST images) by imposing a cut at $i_\mathrm{AB}<25$, resulting in a UV-bright LAE sample of 780 candidates.

	\subsection{INT/WFC}\label{ssection:int}

We use data from the recent  CAlibrating LYMan-$\alpha$ with H$\alpha$ survey \citep[CALYMHA,][]{matthee2016,sobral2017}. This survey aims at detecting LAEs at $z=2.2$ and $z=3.1$ \citep[but also allows the study of other emission lines, see e.g.][]{stroe2017a,stroe2017b}. The observations were made with specially designed filters (NB392, $\lambda_c= 3918\AA$ \& $\Delta\lambda = 52\AA$ and NB501, $\lambda_c= 5008\AA$ \& $\Delta\lambda = 100\AA$) mounted on the Wide Field Camera (WFC) in INT at the Observatorio Roque de los Muchachos on the island of La Palma. To perform the detection of LAEs we use the $U$- and $B$-band images from COSMOS \citep{capak2007} for continuum estimation. Along with the WFC/INT data, we registered the images to the referential frame of HST/ACS survey in COSMOS \citep[][]{scoville2007}. The images were then matched in both spatial resolution ($0.33\arcsec$/pixel) and their Point Spread Function (PSF, which ranges from $1.8-2.0\arcsec$). Fluxes are computed in $3\arcsec$ circular apertures. Candidate Ly$\alpha$ emitters are selected to have rest-frame equivalent widths (EW$_0$) greater than 25\AA\ \citep{sobral2017,matthee2017a}. We perform an additional colour selection aimed at excluding potential interlopers at the redshifts we are probing \citep[see][with respect to NB392 and NB501 colour selections, respectively]{sobral2017,matthee2017a}. We further exclude one additional source based on recent spectroscopic observations (Sobral et al., in prep.). In the end, our WFC/INT sample has a total 159 LAEs at $z=2.2$ and 45 LAEs at $z=3.1$  in the COSMOS region (see Table \ref{tab:summary_bands}).

	\subsection{Subaru/Suprime-Cam}\label{ssection:subaru}

We explore deep data obtained with Subaru Suprime-Cam \citep{miyazaki2002} in the COSMOS field. We have reduced and analysed archival data of 2 narrow-band and 12 medium-band filters that are listed in Table \ref{tab:summary_bands}. The reduction procedure is that described by \citet{matthee2015} and \citet{santos2016}. The extraction of LAEs from the reduced data follows closely the method described in Section \ref{ssection:int} using the appropriate broad band filter data corresponding to each filter for continuum estimation (optical and near-infrared images/catalogues described by \citealt{taniguchi2007,taniguchi2015} and \citealt{capak2007}).  We note that the selection criteria for narrow-band detected LAEs impose a rest-frame equivalent widths EW$_0>25$\AA\ \citep[][]{santos2016}. For medium-band filters, the rest-frame equivalent width cut is at EW$_0>50$\AA. The number of detected LAEs for each processed narrow- and medium-band is shown in Table  \ref{tab:summary_bands}.

	\subsection{HST image data}

High-resolution observations are required to study the rest-frame UV morphological properties of galaxies at high redshift. Thus, we limit our analysis to where HST/ACS F814W images are available \citep[COSMOS survey,][]{scoville2007,koekemoer2007}. We use $10\arcsec\times10\arcsec$ cut-outs of the HST/ACS F814W band images centred on each LAE. These images have a typical PSF FWHM of $\sim$0.09\arcsec, a pixel scale of 0.03$\arcsec/\mathrm{pixel}$ and a limiting point-source depth AB(F814W) = 27.2 (5$\sigma$) and probe the near to far UV for the sources in our sample (on average $\sim$2000\AA\ rest-frame).

Since the LAEs coordinates are measured from narrow/medium band images at a poorer resolution than HST, we are prone to astrometric errors on the centroid estimate. To mitigate this effect we associate to each LAE candidate the closest detected source in the image within 1\arcsec\ radius of the LAE position. If no such source is found, we assume the LAE has no continuum emission in the F814W images (we nevertheless use them in our stacking analysis, see Section \ref{section:method_stacks}).


\section{Methodology}\label{section:methodology}

\begin{figure*}
\centering
\includegraphics[width=\linewidth]{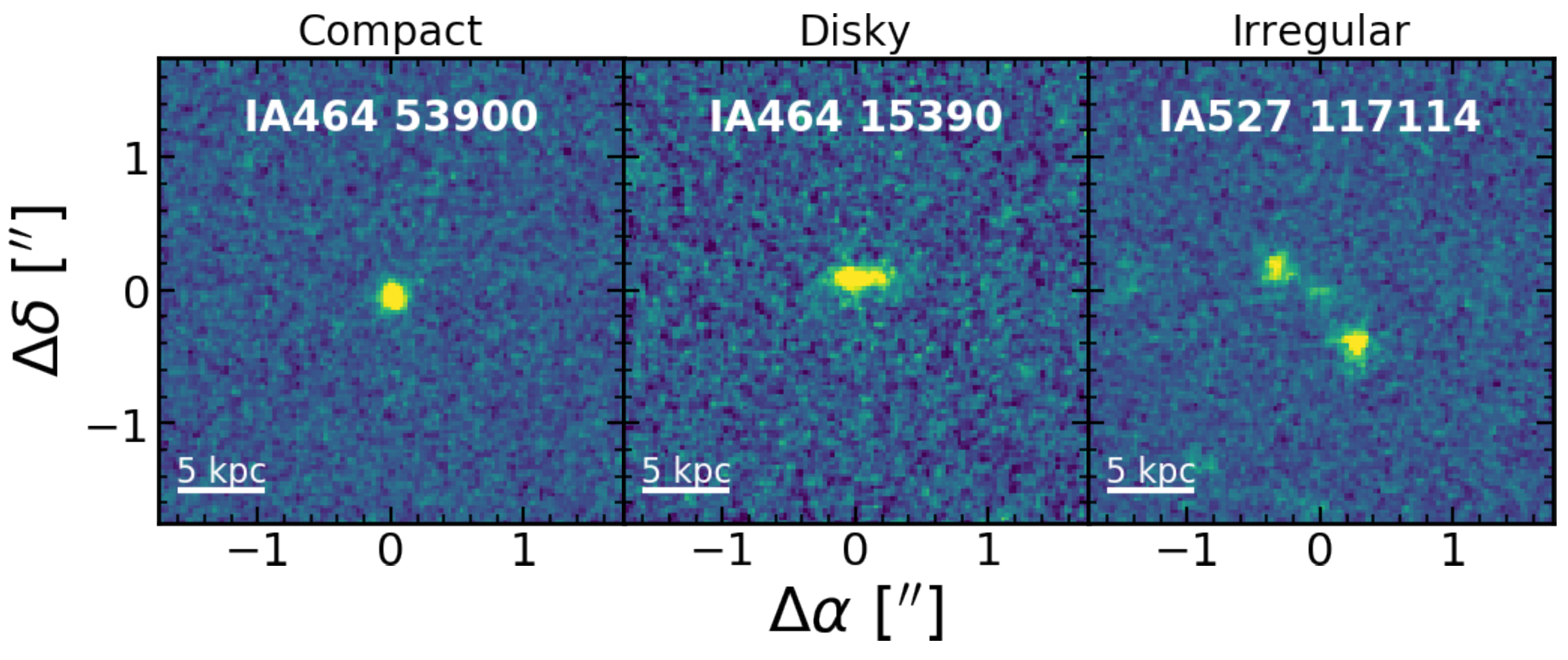}
\caption[Morphological classes of LAEs at $2\lesssim z\lesssim 6$]{Examples of LAEs at $z\sim3$ for each of the morphological classes that we have defined for this study. The classes are displayed from left to right in terms of decreasing compactness: 1) bright round sources with compact profiles, 2) disk-like sources and 3) irregular/mergers/clumpy sources.}
\label{fig:LAEs_visual_examples}
\end{figure*}

To quantify the morphological properties of any given source it is common to fit a parametric model to the observed light profile. In the particular case of galaxy modelling, the \citet{sersic1968} profile is the most common model assumed \citep[e.g.][]{davies1988,caon1993,andreakis1995,moriondo1998,simard1998,khosroshahi2000,graham2001,mollenhoff2001,trujillo2001,peng2002,blanton2003,trujillo2007,wuyts2011,vanderwel2014,shibuya2016} and which is also used to model LAEs \citep[e.g.][]{pirzkal2007,bond2009,gronwall2011}. The S\'ersic model can be described as
\begin{equation}
I (r) = I_e \exp[-\kappa(r/r_{e})^{1/n}+\kappa],
\label{eq:sersic}
\end{equation}
where the S\'ersic index $n$ describes the shape of the light profile, $r_e$ is the effective radius of the profile, $I_e$ is the surface brightness at radius $r=r_e$ and $\kappa$ is a parameter coupled to $n$ such that half of the total flux is enclosed within $r_e$. This profile assumes two characteristic models for specific values of $n$ : exponential disk, if $n=1$, and a \citet{vaucouleurs1948} profile, if $n=4$, best suited for elliptical galaxies and galactic bulges.

An alternative method, relying solely on the observed properties of each object, is to use a non-parametric approach to the morphological characterization \citep[see e.g.][]{abraham1996,bershady2000,conselice2000,conselice2003,lotz2004}. These methods offer reliable estimates even in the case of extremely irregular objects, but fail to account for instrumental effects (such as PSF broadening) and are more susceptible to biases induced by low S/N conditions. 

	\subsection{Structural parameter estimation}\label{ssection:morpgological}
 
The retrieval of structural parameters based on S\'ersic profiles is done using the publicly available \textsc{GALFIT} \citep{peng2002,peng2010}, a stand-alone program aimed at two dimensional decomposition of light profiles through model fitting. In addition to the parameters described in Equation \ref{eq:sersic}\footnotemark{}\footnotetext{In \textsc{GALFIT}, the $I_e$ parameter is computed internally. We use instead the model total magnitude as an input parameter.}, 2D models need 4 additional quantities: the model central position, $x_c$ and $y_c$, the axis ratio of the isophotes, $b/a$ and respective position angle, $\theta_{PA}$, i.e. the angle between the major axis of the ellipse and the vertical axis. 

To run \textsc{GALFIT} effectively, it is necessary that we provide an initial set of parameters. To speed up convergence and minimize the occurrence of unrealistic solutions, it is important that these first guesses provide a good approximation of the light profile. To do so, we use the source extraction software \textsc{SExtractor} \citep{bertin1996}, which can be tuned to produce the parameter set that will be used as input to \textsc{GALFIT}. To fit our galaxies, we use cut-outs centred on each target. The size of the cut-outs was chosen so that we achieve good speed performance and to allow \textsc{GALFIT} to simultaneously fit the residual sky emission. 

To account for the instrumental PSF effects on the observed light profile, we provide PSF images associated with each individual galaxy. We use the HST/ACS PSF profiles that were created with \textsc{TinyTim} \citep{krist1995} models and described by \citet{rhodes2006,rhodes2007}. The PSF model accounts for pixel-to-pixel variation inside the CCD and the different telescope focus value for each COSMOS tile observation. We used the segmentation map produced by \textsc{SExtractor} at the time of the estimation of the initial parameters to create a mask image that flagged all pixels belonging to neighbouring galaxies, preventing them to influence the model of the object of interest. We mask all sources at a distance greater than 1.5\arcsec\ from the target RA, DEC ($\sim$10-13 kpc). We use a morphological dilation (kernel of $3\times3$ pixels) to smooth the individual masked regions and include in the same mask lower flux pixels in the outskirts that are below the \textsc{SExtractor} detection threshold.

Irregular, complex and/or sources detected at low S/N are excluded from the final sample as \textsc{GALFIT} failed to converge on meaningful structural parameters. Note, however, that we also visually classify all sources (see Section \ref{section:method_visual}).

	\subsection{Light concentration}\label{section:method_concentration}

As not all our sources are well fit with a symmetric model ($\sim$45\%), we opted to estimate the light concentration, $C$, of each source by using a non-parametric approach \citep{conselice2000,conselice2003}. We used \textsc{SExtractor} 20\% and 80\% light radius (defined with the parameter \textsc{PHOT\_FLUXFRAC}) and directly computed the value as
\begin{equation}
C= 5 \log_{10}\left(\frac{r_{80}}{r_{20}}\right),
\end{equation}
where $r_{80}$ and $r_{20}$ are the 80\% and 20\% light radius, respectively. This parameter measures the rate of decay of the light profile of galaxies in concentric elliptical apertures and allows us to understand if galaxies have lower or higher surface density of stellar emission in the near- and far-UV. 
Such measure can be linked to the type of star-formation occurring in LAEs which would, in turn, shed some light on the mechanisms linked to the formation of new stars that may boost the escape of Ly$\alpha$ photons. 

	\subsection{Visual classification}\label{section:method_visual}

We complemented the quantification of LAE morphology with the visual classification of the rest-frame UV shapes for all sources with $i_\mathrm{AB}<25$ and HST coverage. We visually classify galaxies in a simple numerical scheme from 0 to 4 in terms of decreasing compactness: 0) corresponding to faint point-like sources, 1) slightly more extended/bright round/not extended sources, 2) disk-like sources and 3) irregular/mergers/clumpy sources (see Figure \ref{fig:LAEs_visual_examples}). Each object was classified independently by three different team members and we combined the final classification by averaging over all classifications. For simplicity, we group classes 0 and 1 as compact sources, 2 as disky and 3 as irregular/clumpy/mergers.

	\subsection{Stacks of Ly$\alpha$ emitters}\label{section:method_stacks}

\begin{figure*}
\centering
\includegraphics[width=\linewidth]{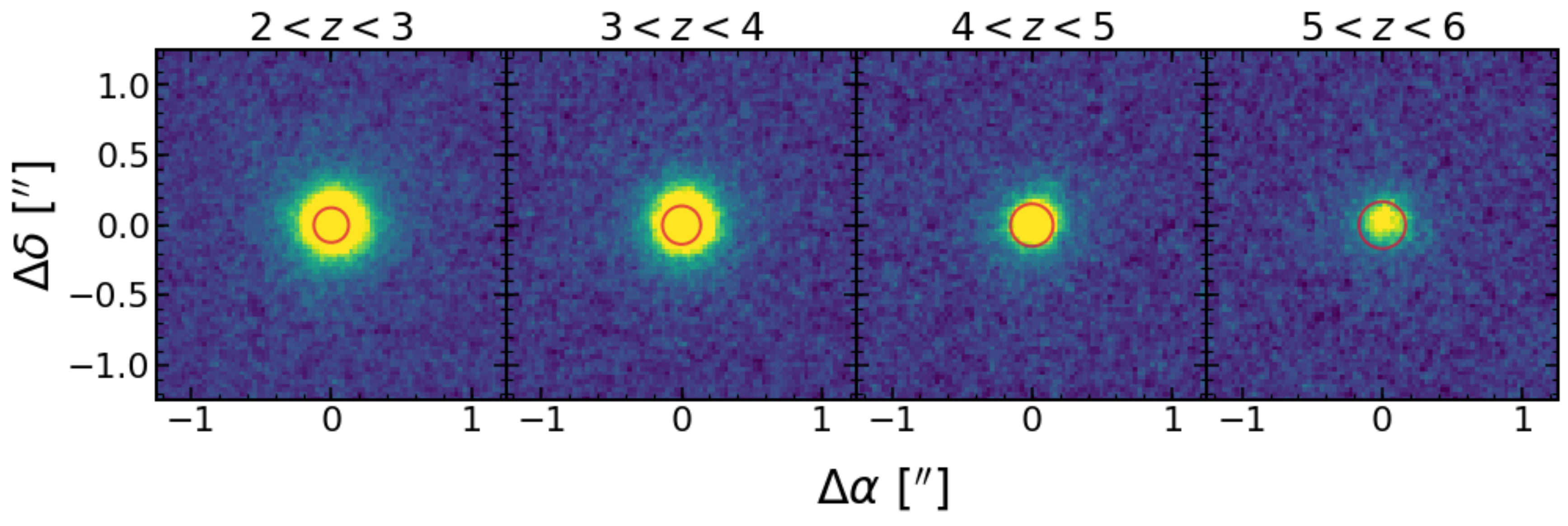}
\caption[LAE image stacks at $2\lesssim z\lesssim 6$]{Examples of LAE stacks for each of the bins that we use in this study in terms of redshift. In each panel, the intensity levels range from -3$\sigma_\mathrm{sky}$ to 15$\sigma_\mathrm{sky}$, where $\sigma_\mathrm{sky}$ is the sky rms. The red circle in each panel has a physical radius of 1 kpc.}
\label{fig:LAEs_stack_examples}
\end{figure*}

The major goal of stacking is to get measurements of the typical galaxy while not being biased by the ones that are brightest in F814W. We have stacked all detected LAEs with available HST/ACS F814W images (a total of 3045 galaxies from all bands) using the median flux per pixel centred at Ly$\alpha$ detection. We have also performed an image shift (typically $\lesssim0.5$\arcsec on the detected sources) since the image coordinates are measured on ground based images and we observe some deviations when seeing them at HST resolution. The resulting stacks, in specific ranges of redshift, Ly$\alpha$ equivalent width and Ly$\alpha$ luminosity are shown in Figure \ref{fig:LAEs_stack_examples} (see also \ref{fig:LAEs_stack_examplesAA}).

We show in Table \ref{tab:summary_bands} the absolute magnitude of these stacks as observed in HST/ACS F814W. These have typical values of $\sim$26 and correspond to absolute magnitudes, in F814W, ranging from $M_\mathrm{i,NB392} = -18.3$ (at low redshift) up to $M_\mathrm{i,IA827} = -20.0$ (at high redshift). These magnitudes are typically 1 to 2 magnitudes lower than $M^\ast_\mathrm{UV}$ at all redshifts \citep[e.g.][]{reddy2009,bouwens2015,finkelstein2015,parsa2016,alavi2016}.

One of the quantities that is affected by the uncertainties on the astrometry of LAEs and a possible mismatch between the peak of Ly$\alpha$ emission and the UV emission \citep[see e.g.][]{shibuya2014a} is the size of the produced light profiles. As we combine astrometric errors from a large number of sources, the profile tends to enlarge. To correct for this, we have used the subset for which we have UV detections in HST ($i_\mathrm{AB}<25$) to compute the difference when using or not a centring algorithm prior to the image stacking. We find that when we do not use a centring algorithm, we produce stacks with an effective radius $\sim$1.1-1.5 times larger. We have computed individual corrections for each of the stacks and morphological quantities $r_e,\ n,\ C$ and applied to all values reported in this work (see e.g. Figure \ref{fig:LAEs_stack_corrections}).


\section{Morphological properties of LAEs}\label{section:results}

In total, we have full individual morphological information on 429 LAEs across $2\lesssim z \lesssim6$ due to GALFIT convergence issues on low S/N galaxies and bright near-perfect point-like objects. However, for visual classification and light concentration parameters, we have results for the 780 LAEs with HST images and $i_\mathrm{AB}<25$. To take full advantage of our large sample, the stacking analysis was done using all 3045 LAEs within the HST footprint. In the next subsections we will detail the rest-frame UV morphological properties of each sample and compare it to the strength of the Ly$\alpha$ emission. We stress that all our results presented in the next subsection are limited to LAEs with $i_\mathrm{AB}\leq25$. For a summary of our findings, see Table \ref{tab:summary_medvalues}. We have excluded X-ray detected AGNs from the sample (see Calhau et al., in prep. for details on AGN selection).

\begin{figure}
\centering
\includegraphics[width=\linewidth]{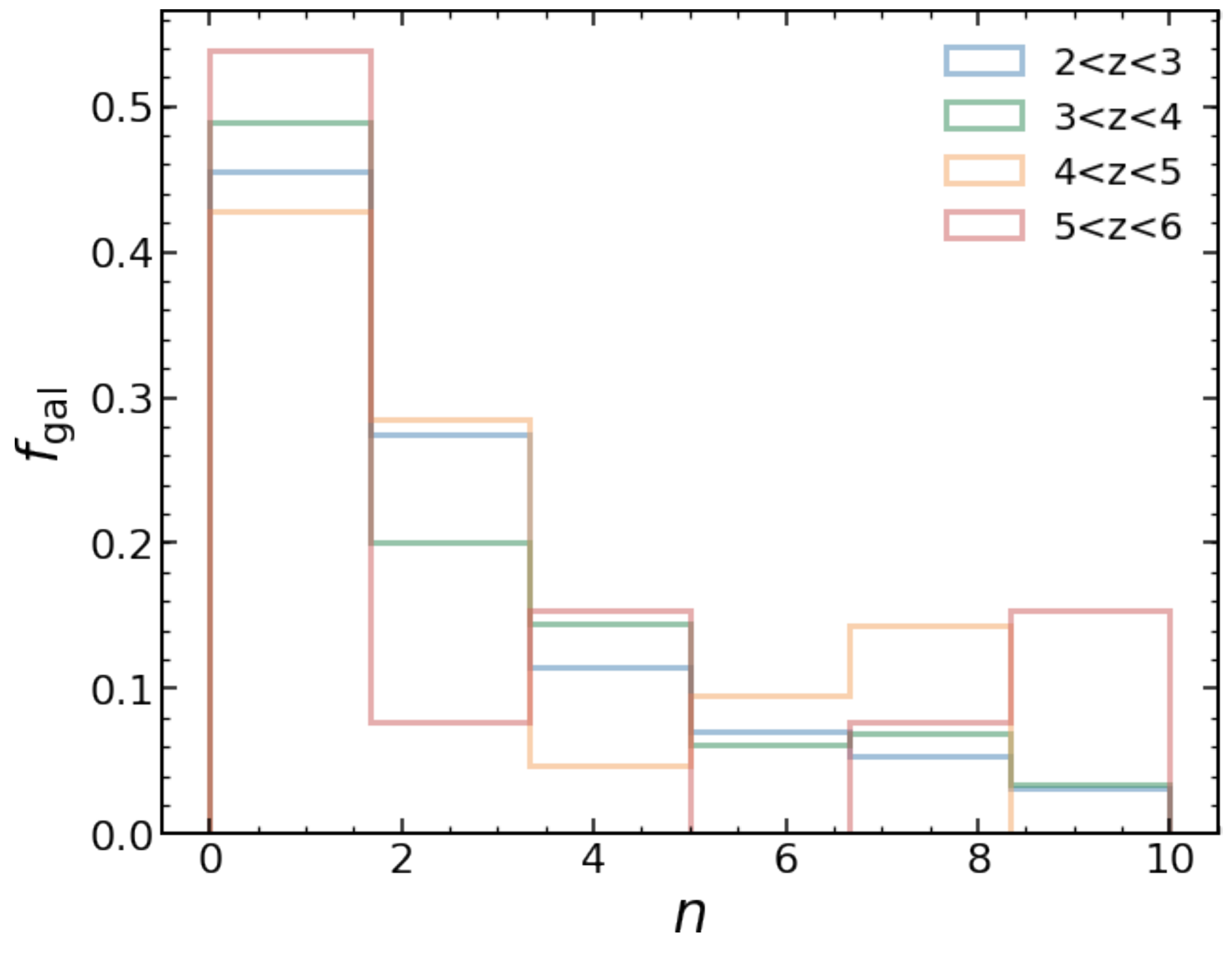}
\includegraphics[width=\linewidth]{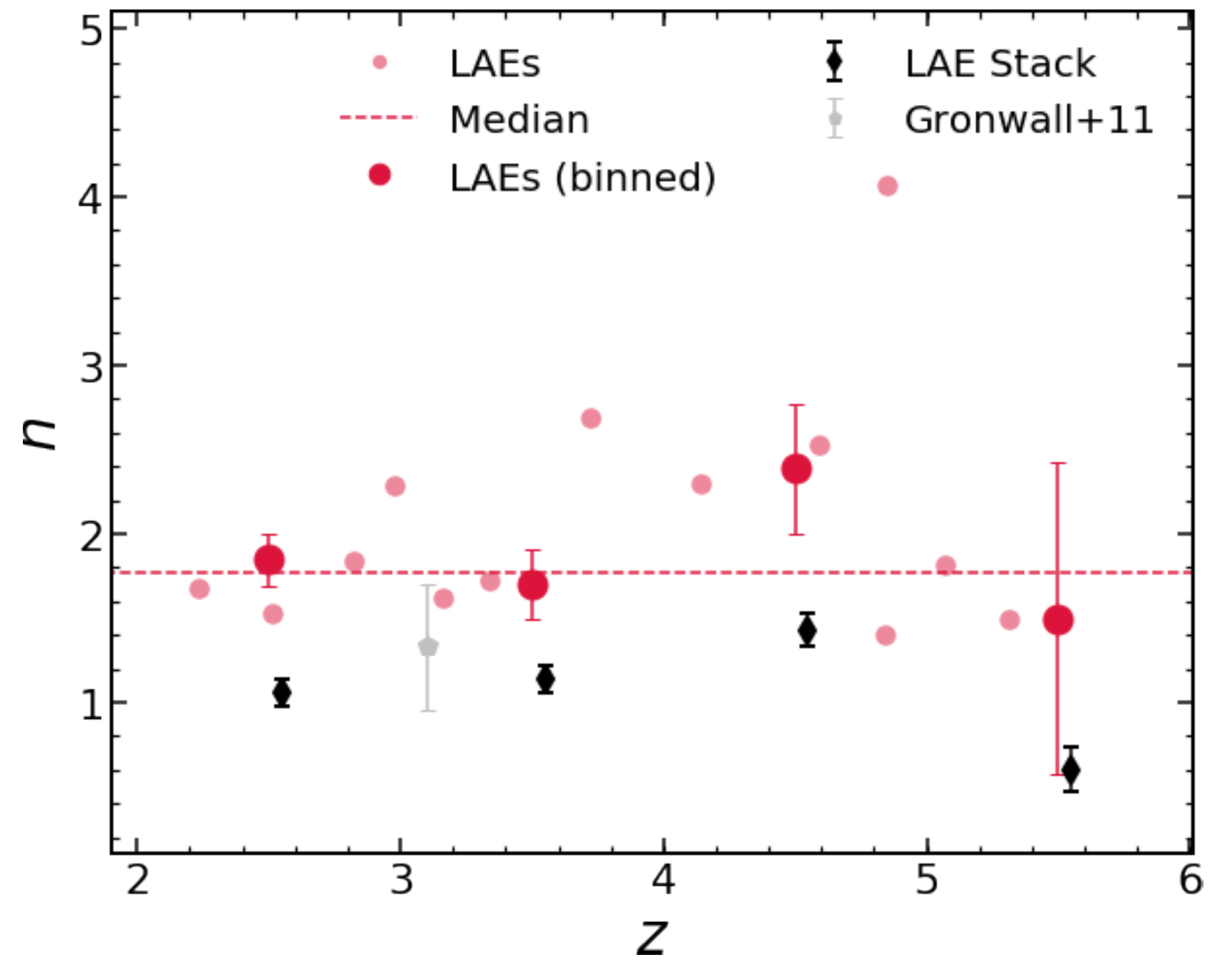}
\caption[S\'ersic index median values of LAEs at $2\lesssim z\lesssim 6$]{S\'ersic index median values of LAEs at $2\lesssim z\lesssim6$. On the top panel we show the S\'ersic index distribution of LAEs for each redshift bin considered. On the bottom panel we plot the evolution of the median S\'ersic index of the distribution (our results in semi-transparent red circles/squares for intermediate/narrow bands and large opaque red circles after binning in redshift) and compare our values to those reported by \citet[][\!\!, grey pentagon]{gronwall2011}. The red dashed line marks the median S\'ersic index for individual LAEs at any redshift. The black diamonds show the S\'ersic index of the stacked LAEs.}
\label{fig:LAEs_sersic}
\end{figure}

\begin{figure}
\centering
\includegraphics[width=\linewidth]{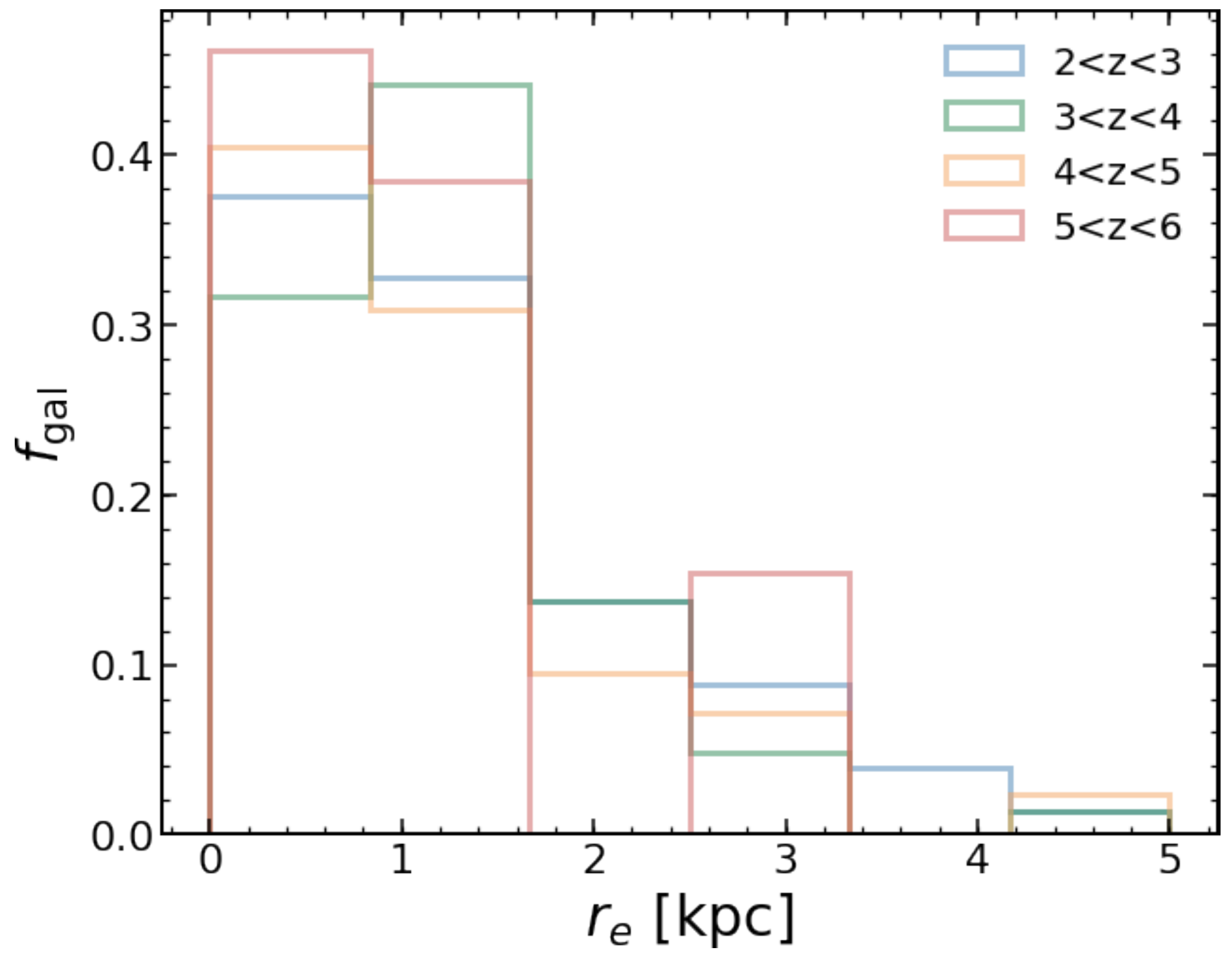}
\includegraphics[width=\linewidth]{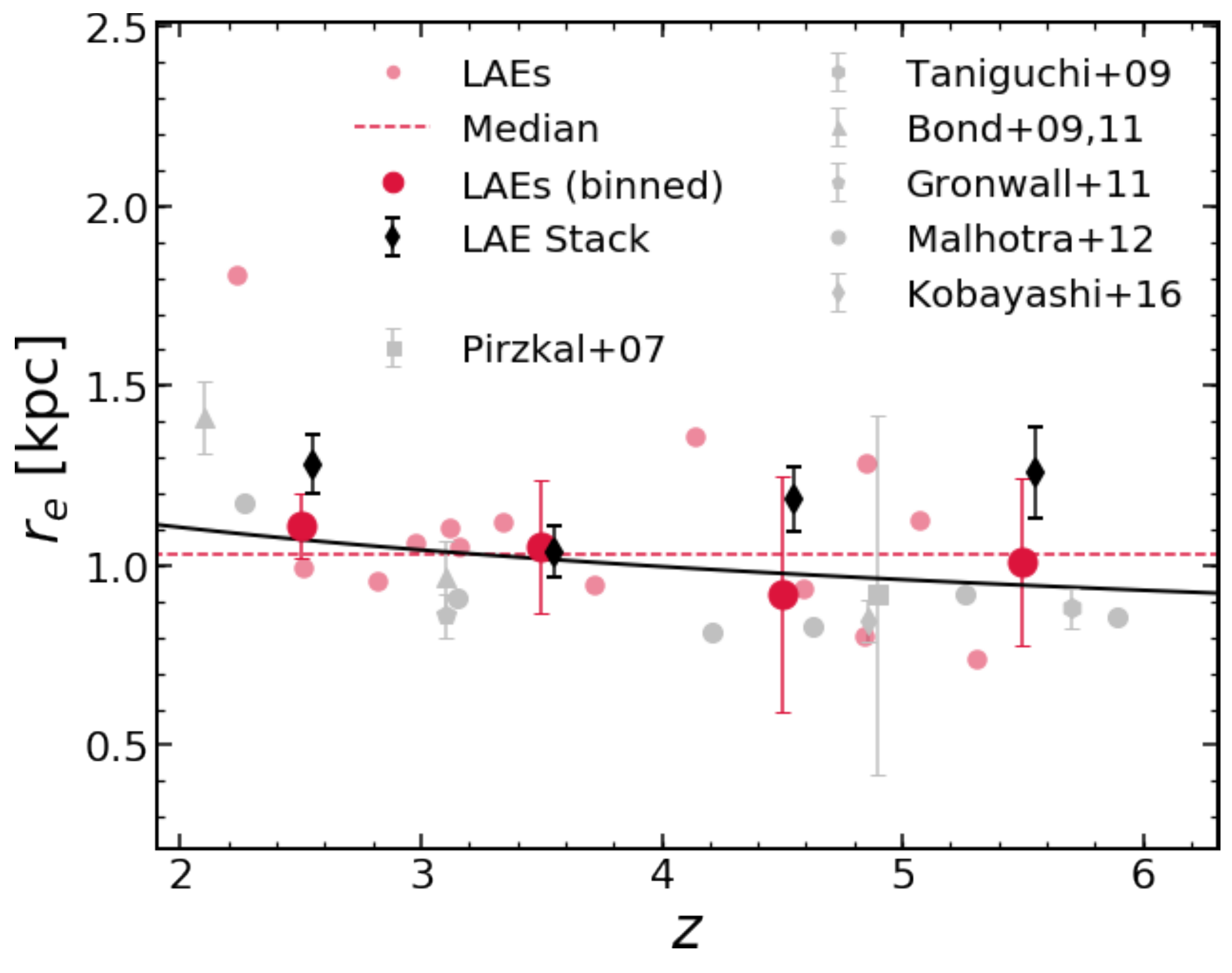}
\caption[Size properties of LAEs at $2\lesssim z\lesssim 6$]{Size properties of LAEs at $2\lesssim z\lesssim6$. On the top panel we show the size distribution of LAEs for each redshift bin considered. On the bottom panel we plot the evolution of the median size of the distribution (our results in semi-transparent red circles/squares for intermediate/narrow bands and large opaque red circles after binning in redshift) and compare our values to those reported in the literature (in grey): square \citep{pirzkal2007}, hexagon \citep{taniguchi2009}, triangles \citep{bond2009,bond2011}, pentagon \citep{gronwall2011}, circles \citep{malhotra2012} and diamond \citep{kobayashi2016}. The black solid line shows the best fit of $r_e \propto (1+z)^{\alpha}$. The red dashed line marks the median effective radius for individual LAEs at any redshift. The black diamonds show the effective radius of the stacked LAEs.}
\label{fig:LAEs_sizes}
\end{figure}

	\subsection{S\'ersic indices and sizes}\label{ssection:ss}

In Figure \ref{fig:LAEs_sersic} we show the distribution of S\'ersic indices and corresponding median of the population. It is readily noticeable that most LAEs have disk-like profiles ($n<1.5$) with fractions ranging from $\sim39\%$ at $4<z<5$ up to $\sim54\%$ at $5<z<6$ (at $2<z<3$ and $3<z<4$ the fractions are of 45\% and 49\%, respectively). We note that there are LAEs with high values of the S\'ersic index. Such cases can be related to galaxies with evident interactions, asymmetric morphologies or compact spheroidal object which we can check with our visual morphological classification. We parametrize the redshift evolution as
\begin{equation}
X = \beta (1+z)^{\alpha}
\label{eq:redshift_evol}
\end{equation}
with $\alpha,\beta$ being the parameters to be fit and $X$ the dependent variable, $n$ in this case. We find that $n\propto (1+z)^{-0.78\pm0.71}$ for the median of the LAE population, which is consistent (at the $\sim1\sigma$ level) with a scenario of no evolution in the light profiles of LAEs. We find that for the lower redshift bins, our reported median values for the S\'ersic index are in good agreement with those reported by \citet{gronwall2011}. We find systematically lower S\'ersic indices for measurements of stacks of LAEs than for individual detections. These difference are related to the smoothing of the central region of the light profile caused by uncertainties on the astrometry (in random directions) and Ly$\alpha$-UV offset which dilute the light and make the profile shallower. Nonetheless, the reported trend is also consistent with little evolution with redshift. 

We show in Figure \ref{fig:LAEs_sizes} the overall properties of the LAE population in 4 bins spanning the redshift range $2\leq z \leq6$. One of the first results is that LAEs have similar size distributions at all redshifts, with most galaxies having effective radii smaller than 1.5 kpc and with $\sim20\%$ as extended sources with $r_e$ from 2-5 kpc. This similarity extends to the evolution on the median population sizes from $z\sim6$ to $z\sim2$, where we observe that LAEs are consistent with little to no evolution scenario in terms of their extent. These results are in agreement with previous results in the literature based on narrow-band selected LAEs \citep[see e.g.][]{venemans2005,pirzkal2007,taniguchi2009,bond2009,bond2011,gronwall2011,malhotra2012,kobayashi2016}. For the evolution of effective radius we find that $r_e \propto (1+z)^{-0.21\pm 0.22}$. This roughly translates to a growth by a factor of $\sim1.2\pm0.2$ for LAEs from $z\sim6$ to $z\sim2$ (consistent with no evolution within $1\sigma$), which compares to a factor of $\sim2.3\pm0.15$ for a more general star forming population \citep[see e.g.][]{vanderwel2014,ribeiro2016}.

We find systematically higher values of the effective radius of measurements of stacks of LAEs than for individual detections. We believe that this is in part due to the centring errors mentioned above but for which we have tried to correct. When deriving size evolution from the stacked profiles, we find that $r_e \propto (1+z)^{-0.01\pm 0.25}$ is perfectly consistent with the lack of evolution that we find for the median population evolution.

We have tested the influence of our choice of binning in the derived parameters and we find that when shuffling the bins we get a variation in the slope $\alpha$, which is always smaller than the reported errors.

	\subsection{Ellipticities}\label{ssections:ellipticity}

\begin{figure}
\centering
\includegraphics[width=\linewidth]{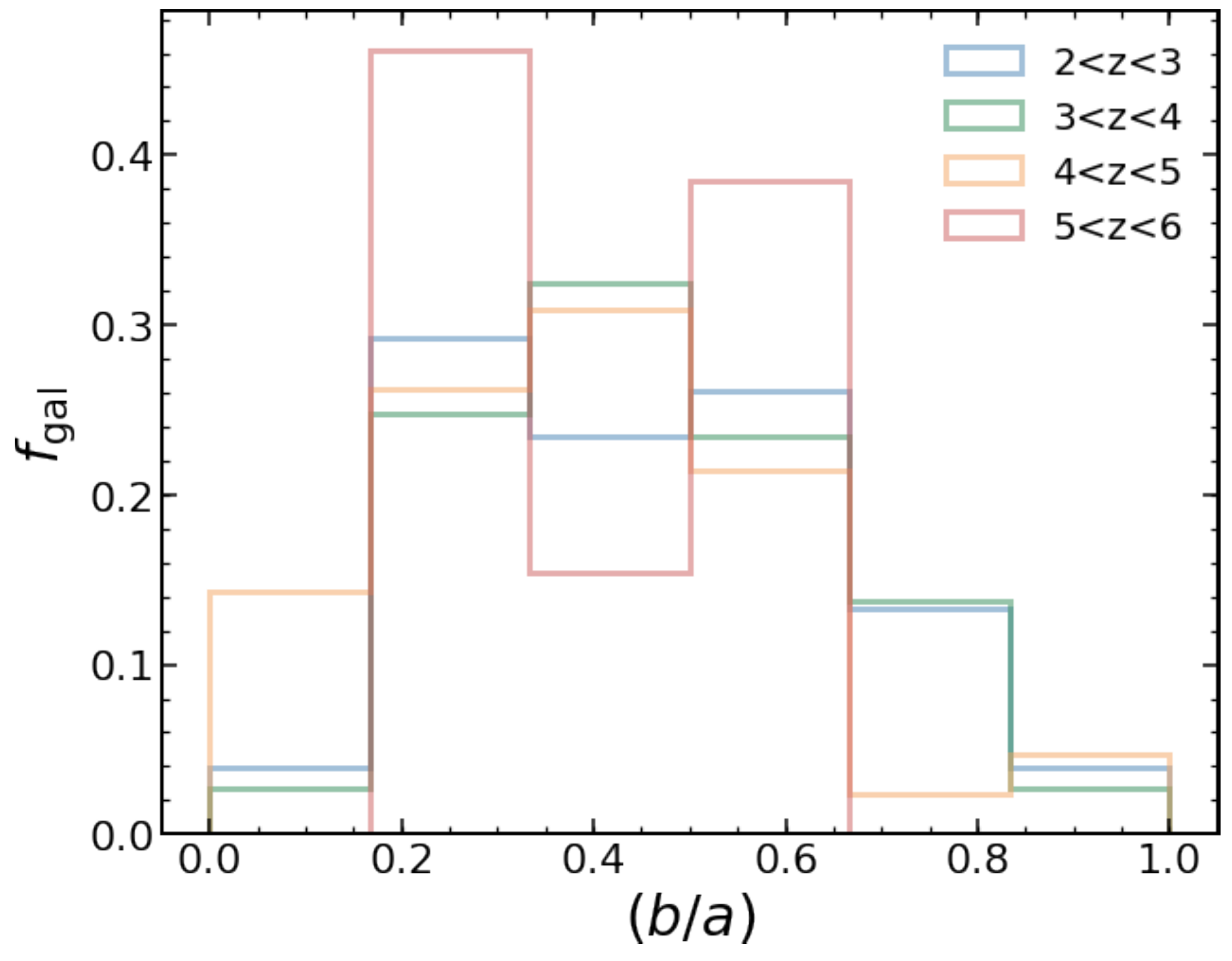}
\includegraphics[width=\linewidth]{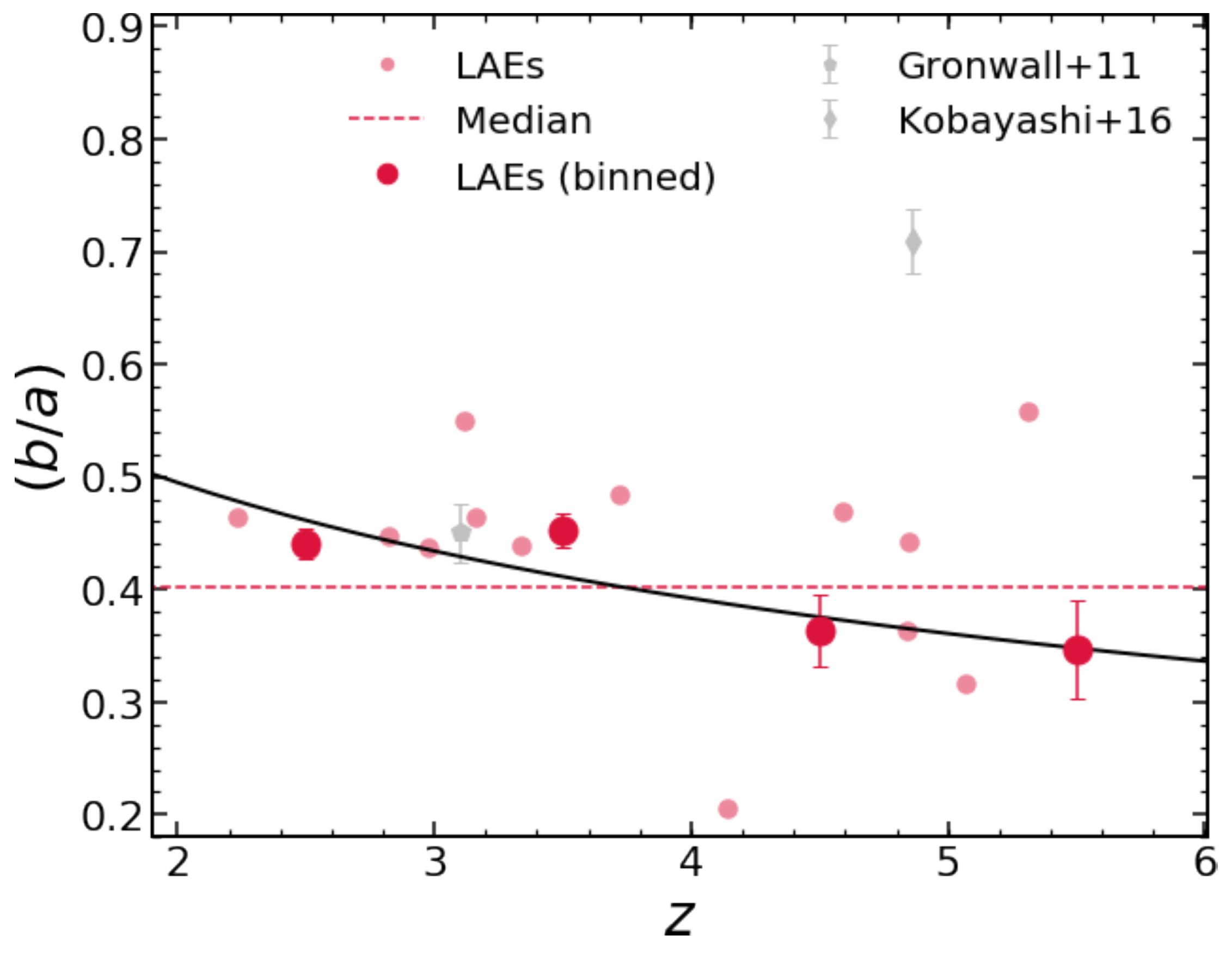}
\caption[Axis-ratio median values of LAEs at $2\lesssim z\lesssim 6$]{Axis-ratio median values of LAEs at $2\lesssim z\lesssim6$. We plot the evolution of the median axis-ratio of the distribution (our results in semi-transparent red circles/squares for intermediate/narrow bands and large opaque red circles after binning in redshift) and we compare our values to those reported in the literature (in grey): pentagon \citep{gronwall2011} and diamond \citep{kobayashi2016}. The black solid line shows the best fit of $(b/a) \propto (1+z)^{\alpha}$. The red dashed line marks the median axis-ratio for all LAEs at any redshift.}
\label{fig:LAEs_q}
\end{figure}

The ellipticity of a source is defined as $e = 1-(b/a)$. We show in Figure \ref{fig:LAEs_q} the results for the derived axis-ratio for the sources in our sample. Our ellipticities have an average uncertainty of 0.07. This is possible due to the use of the HST PSF to convolve with the S\'ersic model before comparing to the data when running GALFIT. We find that LAEs have no clear preference for an ellipticity value, with most of our sources lying at intermediate values $0.2<(b/a)<0.8$. This implies that the detected LAEs do not have to be of a particular shape, which is expected given the randomness of the line-of-sight alignments that determine the 2D shape of each galaxy when viewed through an image. On a more interesting note, this also tells us that a specific alignment of the source with our line-of-sight is not required for it to be detected as a Ly$\alpha$ emitter. These results are in good agreement with measurements at $3<z<5$ by \citet{gronwall2011}. Given the constant S\'ersic indices and the small sizes, our results thus hint that the high Ly$\alpha$ escape fractions of our sources are more of a consequence of their sizes and not orientation effects.

In terms of the median population values for this quantity and its evolution with redshift, we show in the bottom panel of Figure \ref{fig:LAEs_q} that the values of $(b/a)$ are slightly rising with redshift (median value of $(b/a)=0.40$) and in excellent agreement with those reported by \citet{gronwall2011}. However, there is a large discrepancy with the mean value reported by \citet{kobayashi2016} at $z\sim4.86$. We believe that this difference is mostly due to the method used, as they use \textsc{SExtractor} to measure ellipticities that does not account for any PSF broadening which in the case of small galaxies, such as is typical of LAEs, it is natural that the shape is dominated by the PSF in its core, artificially lowering the ellipticity. Using the parametrization of Equation \ref{eq:redshift_evol} we find that $b/a\propto (1+z)^{-0.46\pm0.16}$, which is marginally consistent with a constant ellipticity scenario (within $3\sigma$). This mild evolution reinforces the idea that the galaxy orientation is not a main factor in driving the escape fraction for LAEs.

	\subsection{Concentration}\label{ssections:concentration}

\begin{figure}
\centering
\includegraphics[width=\linewidth]{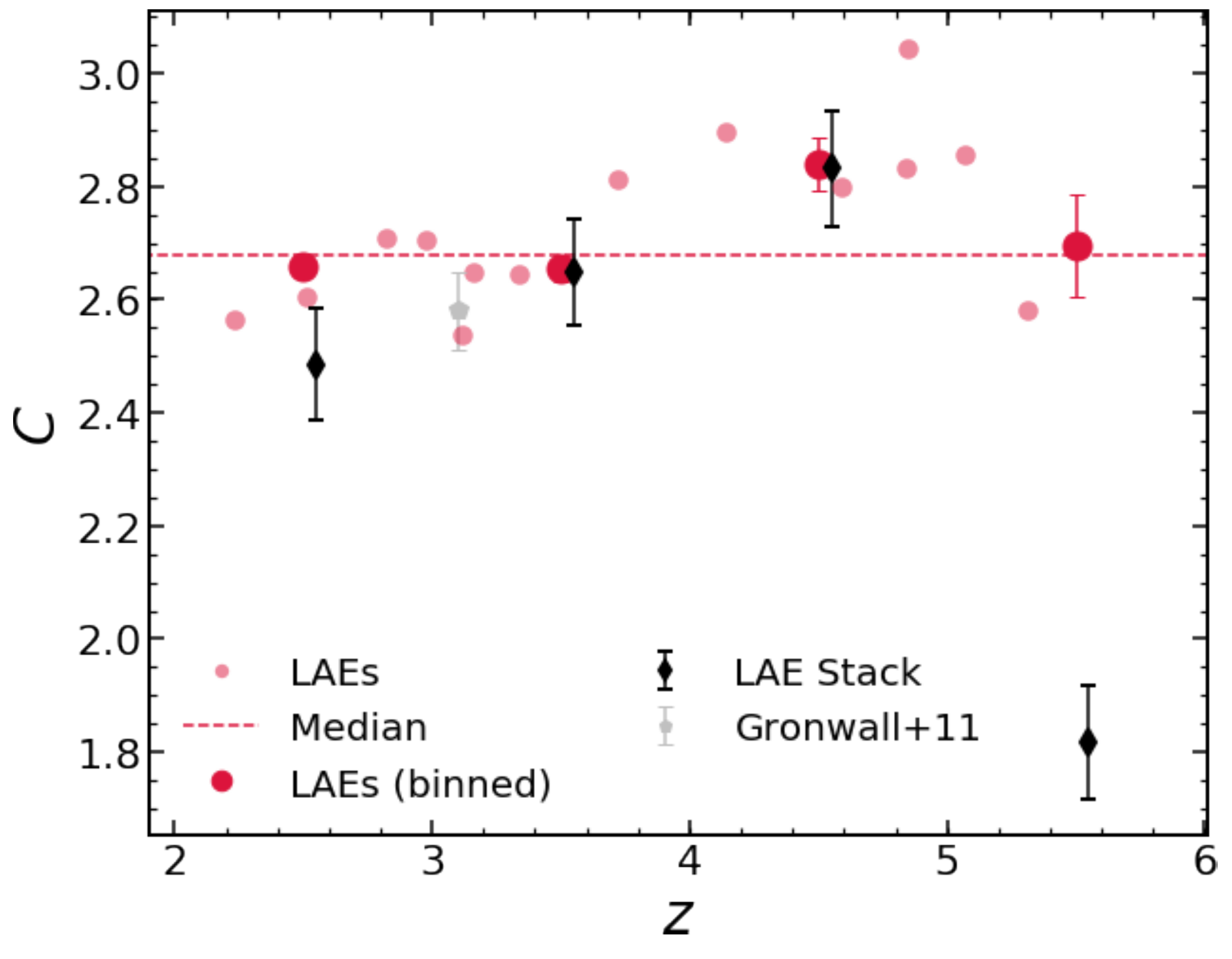}
\caption[Concentration median values of LAEs at $2\lesssim z\lesssim 6$]{Concentration median values of LAEs at $2\lesssim z\lesssim6$. We plot the evolution of the median concentration of the distribution (our results in semi-transparent red circles/squares for intermediate/narrow bands and large opaque red circles after binning in redshift) and compare our values to those reported by \citet[][\!\!, grey pentagon]{gronwall2011}. The red dashed line marks the median concentration for individual LAEs at any redshift. The black diamonds show the concentration of the stacked LAEs. The higher value of concentration for the stacked profiles is likely linked to the combination of the compact nature of these objects and the stacking method we use (see Section \ref{ssections:concentration} for more details).}
\label{fig:LAEs_C}
\end{figure}

In Figure \ref{fig:LAEs_C} we investigate any evolution in terms of the light concentration of galaxies. It is rather stable at $C\sim2.7$ with the exception of the value at $4<z<5$. The fact that this parameter is strikingly similar, in its median evolution, with the S\'ersic index is a possible indication that the galaxies we are probing are rather symmetrical in nature. Both parameters provide a measure of the surface brightness concentration and, in the case of a symmetrical S\'ersic profile, it can be shown that $C$ has a monotonic relation with $n$ \citep[e.g.][]{graham2005}. We find that our results are also in good agreement with the findings by \citet{gronwall2011}. Using the parametrization of Equation \ref{eq:redshift_evol} we find that $C\propto (1+z)^{0.04\pm0.09}$ is fully consistent with a constant light concentration across the entire redshift range. We observe a rise in light concentration for sources at $z\sim4-5$, which is possibly related to an increase on the number of irregular galaxies that we observe. We note that the value at $4<z<5$ is also potentially related to a shallower depth of the images for detection of Ly$\alpha$ (NB711 and IA709), which are more likely to pick sources with higher surface densities and thus higher values of $C$ are to be expected.

The values we find for the concentration of the stacked profiles are consistent with those we find for the median of the population. At the highest redshift, we find much lower concentrations which is potentially related to the higher number of undetected sources that populate this bin allied to the fact that this is also the bin with the fewer galaxies in the stack.

\begin{table*}
\centering
\caption{Median population and stack values as a function of redshift for the morphological quantities presented in this work.}
\begin{tabular}{cccccc}
\hline
& variable & $2<z<3$ & $3<z<4$ & $4<z<5$ & $5<z<6$ \\
\hline
\multirow{4}{*}{Population} & $r_e$ [kpc] & $1.1 \pm 0.1$  & $1.1 \pm 0.2$  & $0.9 \pm 0.3$  & $1.0 \pm 0.2$ \\
& $n$ & $1.8 \pm 0.2$  & $1.7 \pm 0.2$  & $2.5 \pm 0.4$  & $1.5 \pm 0.9$  \\
& $(b/a)$ & $0.44 \pm 0.01$  & $0.45 \pm 0.02$  & $0.36 \pm 0.03$  & $0.35 \pm 0.04$ \\
& $C$ & $2.66 \pm 0.02$  & $2.66 \pm 0.02$  & $2.86 \pm 0.05$  & $2.69 \pm 0.09$ \\
\hline
\multirow{3}{*}{Stack} & $r_e$ [kpc] & $1.2 \pm 0.1$  & $1.0 \pm 0.1$  & $1.1 \pm 0.1$  & $1.3 \pm 0.1$   \\
& $n$ & $1.0 \pm 0.1$  & $1.1 \pm 0.1$  & $1.3 \pm 0.1$  & $0.6 \pm 0.1$   \\
& $C$ & $2.5 \pm 0.1$  & $2.6 \pm 0.1$  & $2.7 \pm 0.1$  & $1.8 \pm 0.1$  \\
\hline
\end{tabular}
\label{tab:summary_medvalues}
\end{table*}

	\subsection{Morphological classes}\label{ssections:morph_class}

\begin{figure}
\centering
\includegraphics[width=\linewidth]{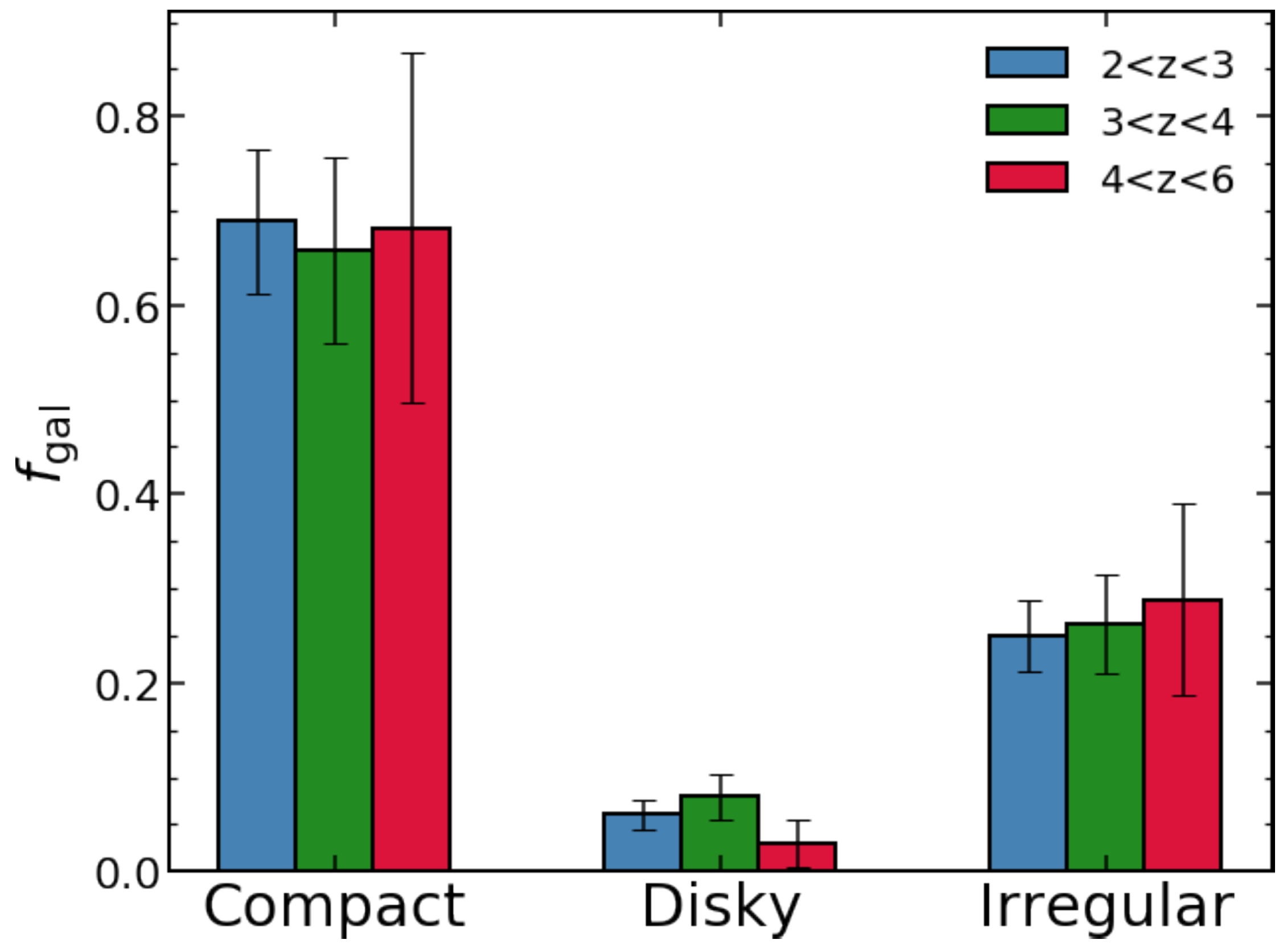}
\caption[Morphological classes from visual classification of LAEs at $2\lesssim z\lesssim 6$]{Fraction of galaxies in each of the morphology classes from visual classification of LAEs at $2\lesssim z\lesssim 6$.}
\label{fig:LAEs_visual}
\end{figure}

Of the 1092 galaxies with $i_\mathrm{AB}<25$ only 780 had good quality images from the HST/ACS archive available. We summarize in Figure \ref{fig:LAEs_visual} the resulting distribution in terms of their visual classification. We find that the majority of our bright LAEs ($\sim67\%$) are found to be compact (point-like+elliptical class). Of the other classes, we find that irregular LAEs are $\sim26\%$ of our sample while disky galaxies amount to only $\sim7\%$ of the observed LAEs. These fractions are roughly constant, but we observe only a slight rise in the fraction of irregulars towards higher redshifts which can be expected of young galaxies in the earlier Universe \citep[e.g.][]{buitrago2013,jiang2013,huertas-company2015b,bowler2017}.

\begin{figure*}
\centering
\includegraphics[width=\linewidth]{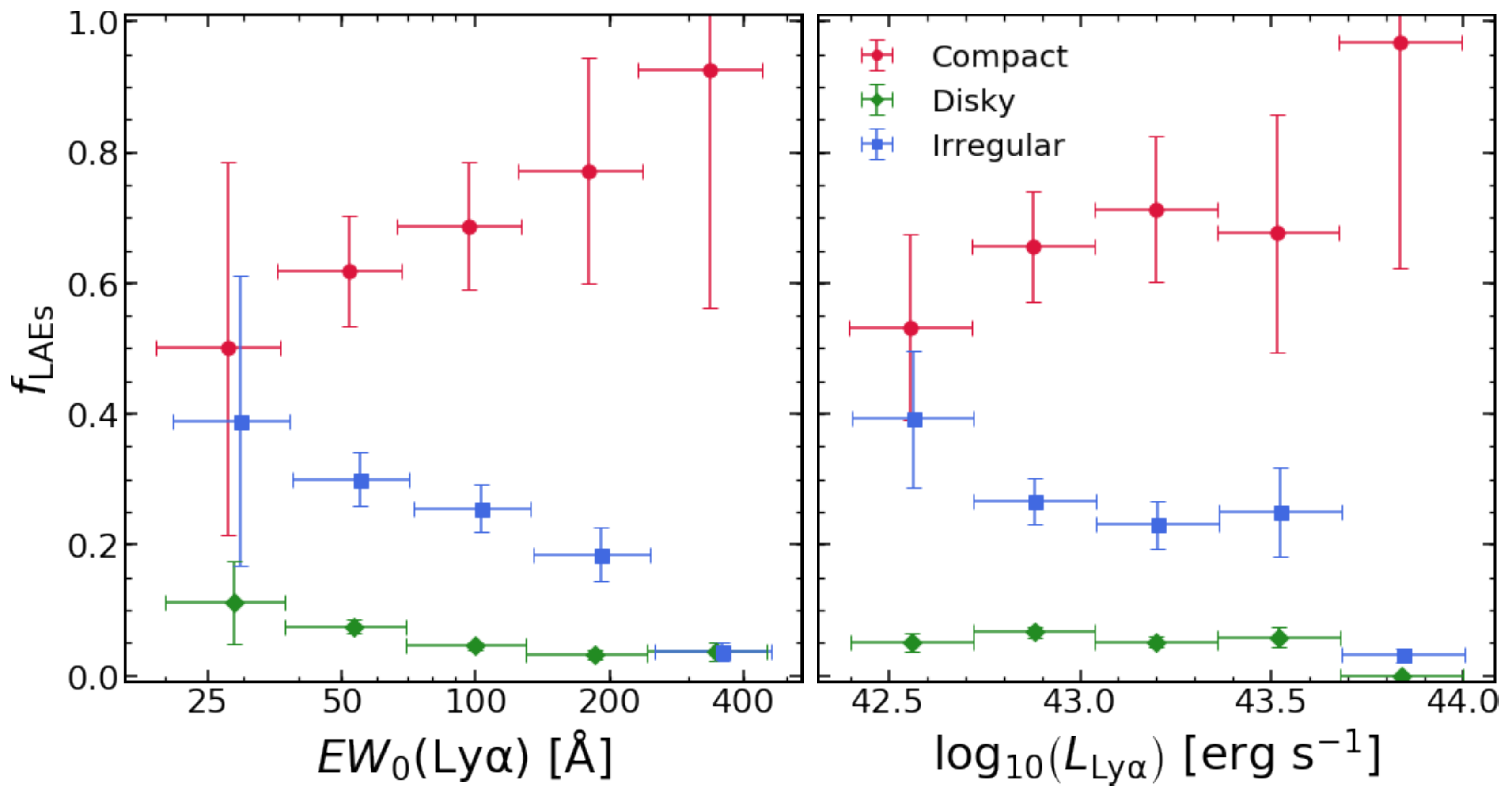}
\caption[Morphological class of LAEs at $2\lesssim z\lesssim 6$]{Fraction of LAEs of a given morphological class (compact as red circles, disky as green diamonds and irregulars as blue squares) at $2\lesssim z\lesssim 6$ as a function of line equivalent width (left) and line luminosity (right).}
\label{fig:LAEs_visual_line}
\end{figure*}

	\subsection{The lack of evolution in individual LAE morphologies}

We have shown in the previous sections the general properties of LAEs in the sample that we are studying and find that the morphology of this population of galaxies is rather stable in this $\sim$3 Gyr period. Since we find that there is not any strong evident evolution in all presented parameters, we opt to study the dependence of Ly$\alpha$ emission properties on the rest-frame UV morphology using the entire sample without discriminating between redshifts (with the majority of our sources being at $z\sim2-3$). This hypothesis will boost the number of sources to inspect such relations and thus uncover more effectively any underlying correlations that may exist. 

We are aware that our sample selection is not done in any absolute quantities (such as in Ly$\alpha$ luminosity or $M_\mathrm{UV}$) and thus we may introduce some biases in our interpretation of the redshift evolution of the presented quantities. We have tested our hypothesis of selection by comparing our results using selections on $\log_{10}(L_\mathrm{Ly\alpha})>43$ and $M_\mathrm{I}<-20.5$, ($M_\mathrm{I}$ is the absolute magnitude of the observed I band) independently. We can report that the lack of evolution in the reported morphological quantities is observed in these smaller subsets from our main sample thus we opt to keep the apparent magnitude cut as our main selection.

	\subsection{Morphology dependence on Ly$\alpha$ luminosity and equivalent width}\label{ssection:geometry}

After summarizing our findings on the morphological properties across cosmic time for the LAEs in our sample, we now turn to the influence of morphology on the observed properties of the Ly$\alpha$ emission itself (line equivalent width and luminosity). 

We show in Figure \ref{fig:LAEs_visual_line} the fraction of each morphological class as a function of line equivalent width and line luminosity. We find that we have no disky galaxies at the lowest equivalent widths and that the irregular galaxies are less common at higher equivalent widths. These trends are accompanied by a slight rise in the fraction of compact galaxies with line equivalent width. We also find that the brightest emitters are tendentiously more likely to be compact than their lower luminosity counterparts. We observe a decline in the fraction of irregular galaxies with line luminosity and a rather stable fraction of disky galaxies at all luminosities that we are probing.

\begin{figure*}
\centering
\includegraphics[height=0.92\textheight]{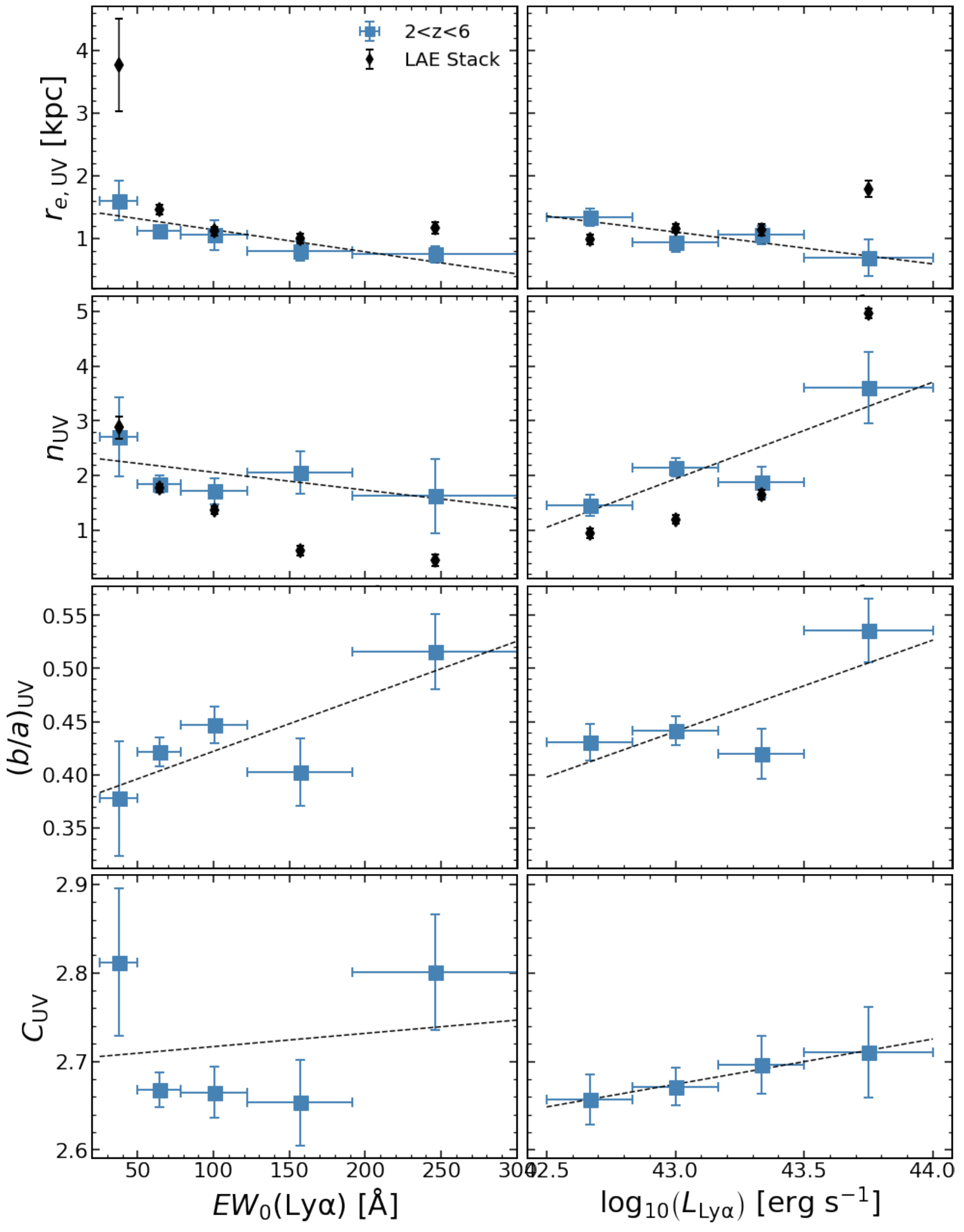}
\caption[Morphology as a function of line equivalent width and line luminosity of LAEs at $2\lesssim z\lesssim6$]{Morphology as a function of line equivalent width (left column) and line luminosity (right column) of LAEs at $2\lesssim z\lesssim6$. From top to bottom we show the results for galaxy size, S\'ersic index, axis-ratio and light concentration. The median values are shown as blue squares and the stacked LAEs properties are represented by the black diamonds. We note that for galaxies with $25\AA < EW_0\mathrm({Ly\alpha})<50\AA$ we only have LAEs detected in the narrow-band surveys, which severely impacts our statistics and give us higher uncertainties in that bin. Black dashed lines show the best linear fits, which have their best parameters shown in Table \ref{tab:summary_fits}.}
\label{fig:LAEs_morph_phys_full}
\end{figure*}

Our results on the relations between morphological quantities and Ly$\alpha$ emission properties are summarized in Figure \ref{fig:LAEs_morph_phys_full} and in Table \ref{tab:summary_fits}.

In Figure \ref{fig:LAEs_morph_phys_full} (first panel, left column), we show the dependence of the equivalent width of Ly$\alpha$ on the observed extent of the UV emission. We observe a trend where higher equivalent width LAEs tend to have smaller sizes. This sort of correlation is seen in other studies \citep[see e.g.][]{taniguchi2009,law2012b,kobayashi2016}, where they find that there is a lack of large galaxies with large equivalent widths and thus the median sizes are naturally smaller at higher equivalent widths. For low equivalent width galaxies, the dispersion on galaxy sizes is larger, spanning the entire interval of measured sizes in the sample. Since Ly$\alpha$ emitters are selected typically with an EW cut then this is likely related with the small sizes of LAEs. Interestingly, this relation may be connected to the physics of Ly$\alpha$ escape since studies have shown that Ly$\alpha$ equivalent width traces the Ly$\alpha$ escape fraction \citep[e.g.][]{sobral2017,verhamme2017}. We highlight that the trend is also observed for the stacked profiles. 

\begin{table*}
\centering
\caption{Linear fits shown in Figure \ref{fig:LAEs_morph_phys_full}. Each line represents a morphological quantity and in the second and third columns we show the parameters from the best fit for Ly$\alpha$ equivalent width and luminosity, respectively.}
\begin{tabular}{ccc}
\hline
y-variable & $x = EW_{0}(\mathrm{Ly\alpha})$ & $x = \log_{10}\left(L_\mathrm{Ly\alpha}\right)$ \\
\hline
$r_e$ & $(-3.5\pm1.2)\times10^{-3}x + (1.5\pm0.2)$ & $(-0.51\pm0.2)x + (23.0\pm8.5)$\\
$n$ & $(-3.3\pm2.3)\times10^{-3}x + (2.4\pm0.3)$ & $(1.8\pm0.63)x + (-74.8\pm27.2)$ \\
$(b/a)$ & $(5.2\pm2.2)\times10^{-4}x + (0.4\pm0.05)$ & $(8.5\pm5.6)\times10^{-2}x + (-3.2\pm2.4)$ \\
$C$ & $(1.6\pm5.5)\times10^{-4}x + (2.7\pm0.1)$ & $(5.2\pm0.72)\times10^{-2}x + (0.4\pm0.3)$ \\
\hline
\end{tabular}
\label{tab:summary_fits}
\end{table*}

We plot in Figure \ref{fig:LAEs_morph_phys_full} (second panel, left column) the relation between S\'ersic index and Ly$\alpha$ equivalent width and find that there is a slight sign of a correlation between these two quantities. Our data suggests that at higher equivalent widths ($EW_0(\mathrm{Ly\alpha})>200$\AA) we are more likely to have shallower profiles. This trend is also seen from the stacked profiles albeit at systematically lower values of $n$ (see Section \ref{ssection:ss}).

A stronger correlation that we find is between the axis-ratio of the emission and the measured line equivalent width. In Figure \ref{fig:LAEs_morph_phys_full} (third panel, left column) we find that the strongest emitters (the ones with the largest equivalent width) tend to have rounder shapes (higher axis-ratios, lower ellipticities). These findings are in agreement with those reported by \citet{kobayashi2016} at $z\sim4.86$. If we assume that the axis-ratio is a good proxy for galaxy inclination, we can explain the observed trend as a simple effect of geometry due to the inclination of the disk with respect to our line of sight \citep[see e.g.][]{verhamme2012,berhens2014}. However, one must be cautious when comparing observations directly with simulations since the latter assume that the galaxy is a perfect flat disk and the former assumes that galaxies are symmetrical enough to be well fit by a parametric model and either assumption has its drawbacks.

We finally explore the correlation between light concentration and Ly$\alpha$ in Figure \ref{fig:LAEs_morph_phys_full} (fourth panel, left column). Much like the case we presented for the S\'ersic index, we cannot infer conclusively about any correlation between these two quantities. We may tentatively say that galaxies with higher equivalent widths ($EW_0(\mathrm{Ly\alpha})>200$\AA) are to be more concentrated in term of the rest-frame UV emission when compared to their lower equivalent width counterparts. The higher concentration value we have for the lower equivalent width bin is explained due to the lower number statistics of that bin. As stated in Section \ref{section:data}, galaxies with $25<EW_0(\mathrm{Ly\alpha})<50$\AA\ are only from narrow band data.

We attempt at a similar exercise as above and explore the possible correlations between the galaxy morphology and its observed Ly$\alpha$ line luminosity. 

Concerning galaxy sizes there is an apparent downward trend for galaxies with $10^{42.5} \lesssim L_\mathrm{Ly\alpha}\lesssim10^{44} \mathrm{erg\ s^{-1}}$ with galaxies being smaller at higher luminosities (see Figure \ref{fig:LAEs_morph_phys_full}, first panel, right column). This trend is not clear since there are some bin-to-bin variations that are mainly due to our small number of objects as well as the loose correlation that exists between these two quantities (for any luminosity bin there is a large spread in galaxy sizes). Interestingly we find an opposite trend when considering the sizes of the stacked profiles. We find this can be explained by an underlying $i_\mathrm{AB}$ - Ly$\alpha$ line luminosity where the brightest galaxies on our sample in rest-frame UV are also the ones with the highest Ly$\alpha$ line luminosity. When stacking a large number of bright galaxies, we are more likely to pick up extended lower surface brightness regions and thus get larger sizes.

We find a similar scenario for the S\'ersic index (see Figure \ref{fig:LAEs_morph_phys_full}, second panel, right column), with higher luminosity LAEs hinting at a higher value of a S\'ersic index. We believe again that the large dispersion on the observed data is likely an indication of the loose correlation between these two quantities. Nevertheless, it is remarkable that the brightest LAEs have such high S\'ersic index ($n\sim3.5$), corresponding to more classical elliptical profiles. This is a consequence of bright, small and compact objects that are more likely to possess such profiles. We find the same response when looking at the values of the stacked LAEs, with high luminosity LAEs ($L_\mathrm{Ly\alpha}\sim10^{43.75} \mathrm{erg\ s^{-1}}$) having higher values of $n\sim4$. We find the same trend when considering the stacked profiles, albeit at a shallower slope and systematically lower values of $n$ (see Section \ref{ssection:ss}).

When estimating the median axis ratio as a function of Ly$\alpha$ line luminosity (see Figure \ref{fig:LAEs_morph_phys_full}, third panel, right column) we find the same trend as compared to the relation with line equivalent width. In this case, galaxies at higher luminosities show less elongated shapes than their lower luminosity counterparts.

Finally, we show in Figure \ref{fig:LAEs_morph_phys_full} (fourth panel, right column) that there is a small but steady increase of the light concentration for our luminosity bins. This trend is less broken that what is reported for the equivalent width of the Ly$\alpha$ line, but still points to a scenario where the brightest Ly$\alpha$ emitters are more likely to have high light concentration in their profiles (as also seen in the S\'ersic index).


\section{The dependence of morphology on Ly$\alpha$}

\begin{figure*}
\centering
\includegraphics[width=\linewidth]{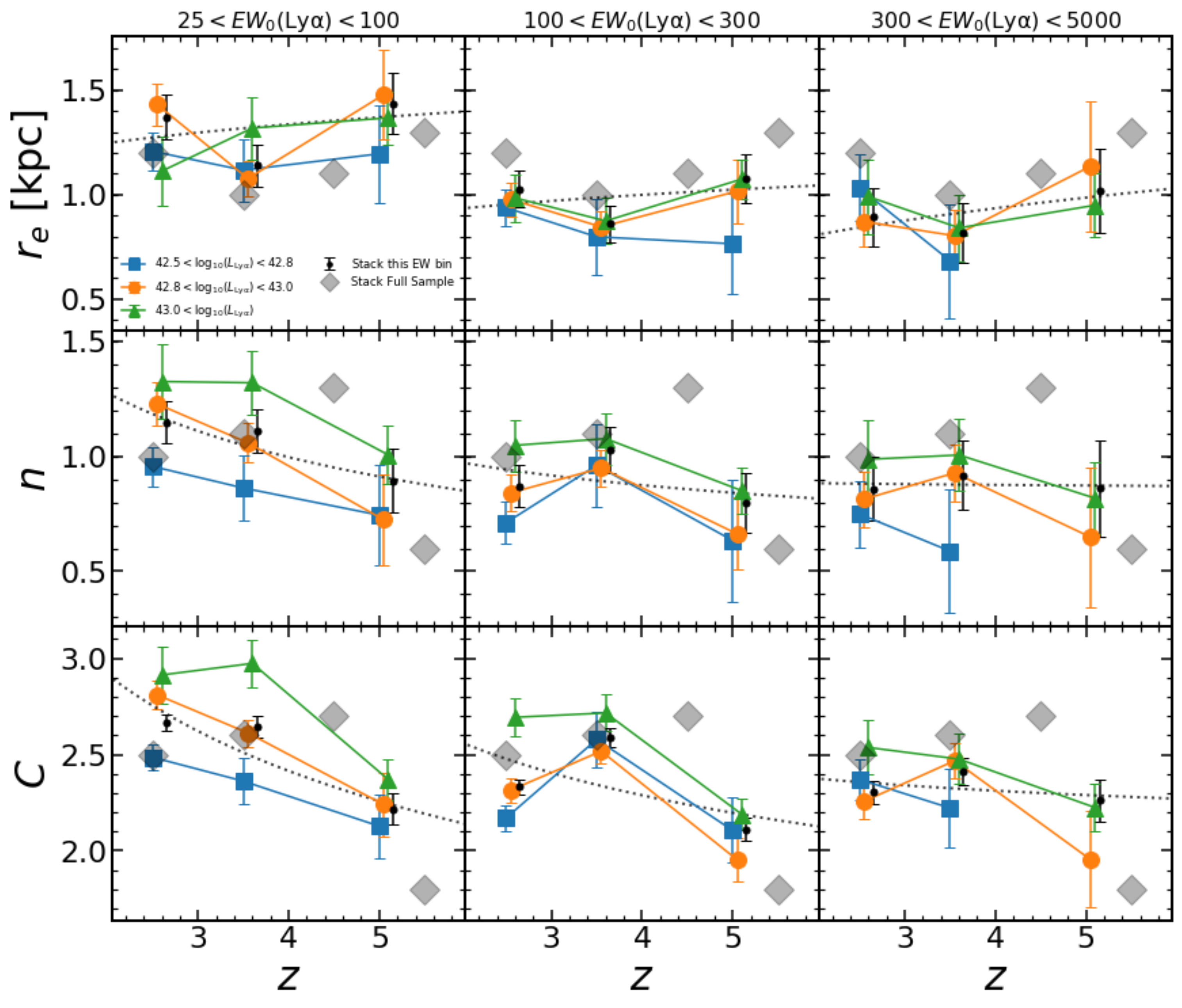}
\caption[Results on individual stacks for multi-parameter exploration]{Morphological evolution results for LAE stacks binned in redshift, Ly$\alpha$ luminosities and Ly$\alpha$ equivalent widths independently. Each row shows a different Ly$\alpha$ equivalent width bin and each line a different Ly$\alpha$ luminosity bin. We also compare those with the stack of the full sample binned in redshift and the stack for a specific EW bin. The dotted black line shows the $X \propto (1+z)^{\alpha}$ fit to the EW bin. Overall we find that galaxy sizes grow with cosmic time for all luminosities and equivalent width bins. We find a similar increase in S\'ersic index and light concentration, from high to low redshift in all redshift bins. We find the strongest evolution in galaxy sizes, S\'ersic index and light concentration in the lower Ly$\alpha$ equivalent width galaxies.}
\label{fig:LAEs_cubeStack_results}
\end{figure*}

To investigate the dependence of UV morphology of LAEs on redshift, Ly$\alpha$ luminosities and  Ly$\alpha$ EWs independently we have stacked different subsamples probing individually each parameter. The resulting stacked images are shown in Figures \ref{fig:LAEs_cubeStack1}-\ref{fig:LAEs_cubeStack3}. We followed the methodology detailed in Section \ref{section:method_stacks} and obtained individual morphological parameters for each subsample. We summarize the results in Figure \ref{fig:LAEs_cubeStack_results}. We did not perform similar analysis on the median of the population since we only have individual measurements on 429 galaxies. Doing so would result in a highly diluted sample over the 27 bins that we need to probe independently the correlations (3 bins per independent variable). We note however that due to scarcity of sources with $i_\mathrm{AB}<25$ we are not able to compute the corrections described in Appendix \ref{section:appendixA}. We then use the average redshift correction we derive and apply it to all Ly$\alpha$ luminosities and  Ly$\alpha$ EW bins.

	\subsection{Galaxy sizes}

When considering the evolution in LAEs sizes we find that independently probing in Ly$\alpha$ luminosities and  Ly$\alpha$ equivalent width yields interesting results. Overall, we find no significant size evolution in any of the bins that we probe. To quantify the evolution trends that we find we fit Equation \ref{eq:redshift_evol} to each EW bin defined. We find that the slopes to be $\alpha=-0.11\pm0.42$ for $25<EW<100$\ \AA\ stacked LAEs, $\alpha=-0.11\pm0.41$ for $100<EW<300$\ \AA\ stacked LAEs and $\alpha=0.27\pm0.31$ for $EW>300$\ \AA\ stacked LAEs. These slopes are all perfectly consistent with no evolution as we have derived when considering all galaxies (see Figure \ref{fig:LAEs_sizes}). In addition to that, galaxies with the lowest Ly$\alpha$ EWs are also the largest at any redshift \citep[similar to what is found in e.g.][]{law2012b}. This is the same trend as we already find in Figure \ref{fig:LAEs_morph_phys_full} when considering the entire LAE population. We find larger galaxies for low EW LAEs.

The luminosity of Ly$\alpha$ seems not to have such a strong influence on the measured sizes (as already indicated by contrary trends in Figure \ref{fig:LAEs_morph_phys_full}). It is true that we have small indications that higher luminosity galaxies can, on average, be larger but that is only marginally true for the intermediate line equivalent width subsamples ($100<EW<300$\ \AA). At lower and higher equivalent widths the picture is not so clear and we get different size-luminosity trends at different redshifts. 

We nevertheless stress that we may be affected by measurement errors on low S/N stacked images, as we have fewer galaxies at high redshift to stack from. We show in Figures \ref{fig:LAEs_cubeStack1}-\ref{fig:LAEs_cubeStack3} that at $4<z<6$ we have faint surface brightness for low and intermediate luminosity bins. We would need larger samples and/or deeper high resolution surveys to probe the morphology of the LAE population.

	\subsection{S\'ersic index and light concentration}

In highly symmetrical profiles, such as those produced by stacking a large number of galaxies with random orientations in the sky, it is normal that the light concentration index, $C$, correlates with the S\'ersic index, $n$. In that sense, the results that we find for both are qualitatively the same and can be interpreted from the same perspective. We find an overall increase of both morphological indexes with redshift. We also observe an increase of both indexes with Ly$\alpha$ luminosity and a decrease with Ly$\alpha$ equivalent width. In summary, we expect that the light concentration would be higher for high Ly$\alpha$ luminosity, low Ly$\alpha$ equivalent width and low redshift and lower light concentration for low Ly$\alpha$ luminosity, high Ly$\alpha$ equivalent width and high redshift.

These results are in good agreement with the more general trends that we find in Figure \ref{fig:LAEs_morph_phys_full}, where we also see similar relations when looking at the whole population. It also reinforces the idea that the trends in Ly$\alpha$ luminosities and Ly$\alpha$ equivalent widths are valid in the redshift bins that we probe here. However, we find more interesting trends concerning the evolution with redshift which are slightly more pronounced when separating the galaxies in different regions of the Ly$\alpha$ luminosity and Ly$\alpha$ equivalent width parameter space.


\section{Discussion}\label{section:discussion}

	\subsection{The evolution in LAE morphology between $z\sim2-6$}

Our results regarding galaxy morphology as a function of redshift (see Section \ref{section:results}) indicate that LAEs have the same typical shape across the period we probe ($z\sim2-6$). This is reflected by the little to no variation in size, S\'ersic index, axis-ratio and light concentration parameters which is seen both in the median of the population as well as in the stacked profiles. This lack of evolution relates only to the general LAE population taken as whole. As we show in Figure \ref{fig:LAEs_cubeStack_results}, we are maybe seeing some redshift evolution in S\'ersic indexes and light concentration when separating our sample in different Ly$\alpha$ luminosity and Ly$\alpha$ EW bins (but no evidence of that for galaxy sizes). The strongest redshift evolution that we find is for low Ly$\alpha$ equivalent width galaxies. This is likely because typical SFGs and/or LBGs with Ly$\alpha$ emission are mostly populated with low Ly$\alpha$ equivalent widths \citep[see e.g.][]{hathi2016} and thus we can expect that these subsamples would have similar properties to that population. We note, however, that the lowest equivalent width LAEs (which are those brighter in UV and likely the most massive within our sample) are not the dominant population and that is the main reason why we do not see strong evolution in morphology of the LAE population from $z\sim2$ to $z\sim6$.

	\subsection{LAE sizes at $z\sim2-6$}\label{ssection:sizeevolution}

\begin{figure*}
\centering
\includegraphics[width=\linewidth]{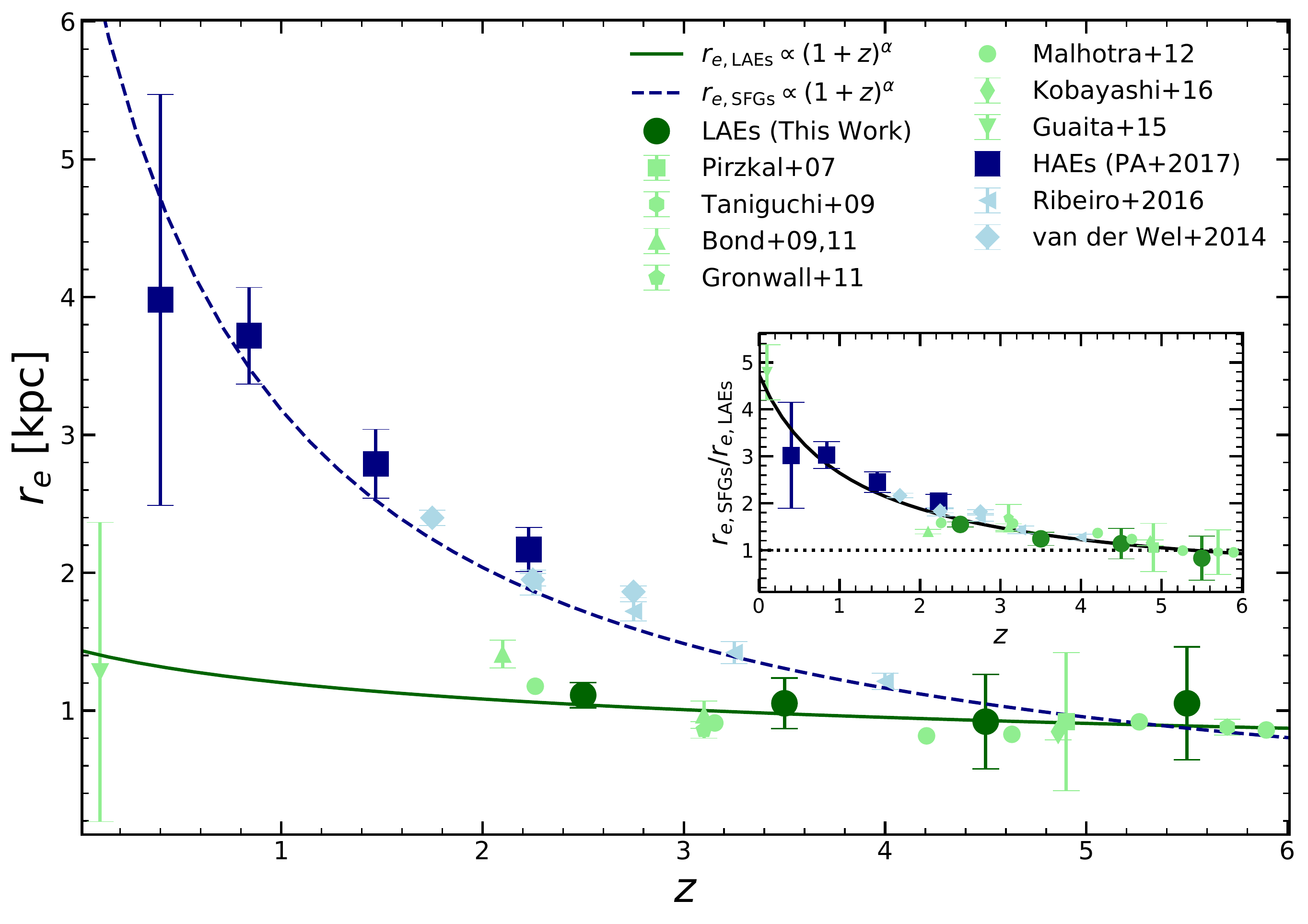}
\caption[Size properties of LAEs at $2\lesssim z\lesssim 6, comparison with HAEs$]{Size properties of LAEs at $2\lesssim z\lesssim6$. We plot the evolution of the median size of the distribution (our results in large green circles) and compare our values to those reported in the literature (in light green): square \citep{pirzkal2007}, hexagon \citep{taniguchi2009}, triangles \citep{bond2009,bond2011}, circles \citep{malhotra2012}, diamond \citep{kobayashi2016} and inverted triangle \citep{guaita2015}. We show as blue squares the median size for a sample of HAEs selected at lower redshift using the same narrow band technique \citep{paulino-afonso2017}. We complement this figure with results for UV-selected star-forming galaxies from the literature (in light blue): large diamond \citep{vanderwel2014} and left-facing triangle \citep{ribeiro2016}. Finally, we show the derived size evolution of LAEs (green solid line) and SFGs (blue dashed line). The inset plot shows the estimated size ratio between SFGs and LAEs. Estimates point to SFGs being $\sim5$ times larger at $z\sim0$ and of the same size as LAEs at $z\sim5.5$. We hypothesize that Ly$\alpha$ selected galaxies are small/compact throughout cosmic time likely linked with the physical processes that drive Ly$\alpha$ escape. At higher and higher redshifts, typical SFGs start to have the typical sizes of Ly$\alpha$ emitters, which can be seen as an alternative explanation for the rise of the Ly$\alpha$ emitting fraction of SFGs/LBGs into $z\sim6$.}
\label{fig:LAEs_sizes_wHAEs}
\end{figure*}

Finally, in Figure \ref{fig:LAEs_sizes_wHAEs} we show our results for the evolution in rest-frame UV sizes of LAEs across cosmic time and compare our findings to previous studies \citep[e.g.][]{taniguchi2009,bond2011,bond2012,gronwall2011,guaita2015,kobayashi2016}. Our median effective radius are in agreement with other size estimates of LAEs in the literature. At $z\gtrsim4$ we find typical sizes of $r_e\sim0.9$ kpc and at $z\sim2.2$ we find slightly larger galaxies with average sizes of $r_e =1.1$ kpc. We have attempted to fit a relation to our data points and find that $r_e \propto (1+z)^{-0.21\pm 0.22}$ (see Section \ref{ssection:ss}). This scenario, however, predicts slightly larger sizes at $z\sim0$ than what have been reported for the LARS sample in the local Universe \citep{guaita2015} but within their reported dispersion. 
This scenario points to a lack of evolution on the sizes of LAEs since $z\sim6$. However, this reasoning hinges on the single point that we have at $z\sim0$ and which is derived from a heterogeneous sample of 14 galaxies only. To fully understand if LAEs evolve in size as hinted by the data at $z\sim2.5$ one would need larger samples between $z=0-2$, which are currently out of the scope of any instrument apart from HST/COS.

	\subsection{Relations between LAEs, HAEs and UV-selected galaxies}\label{ssection:uvsizes}

When compared to the typical sizes of star-forming galaxies (selected as H$\alpha$ emitters, HAEs) that have been studied in a previous work \citep{paulino-afonso2017}, we immediately see that the two populations are not alike in terms of their extent. Despite having only one common period with observations of both populations (at $z=2.23$), where HAEs are almost two times larger than LAEs, our prediction of LAEs sizes at lower redshifts are consistently lower that what we report for the HAE population. This is even more contrasting if we include the sample at $z\sim0$, where almost no evolution is expected for the LAEs population. This is potentially corroborated by the existence of green pea galaxies \citep{cardamone2009,izotov2011} which are compact in nature and found to have Ly$\alpha$ detections and high Ly$\alpha$ escape fractions \citep[e.g.][]{henry2015,yang2016,verhamme2017}.

We also use two recent and comprehensive studies on the evolution of UV-selected star-forming galaxies \citep[][\!\!, see also \citealt{shibuya2015}]{vanderwel2014,ribeiro2016} that overlap both HAEs and LAEs that we have studied to complement our observations. These confirm our findings that at $z\sim2$ the typical star-forming population is larger in size than the LAEs population (by a factor of $\sim3$). However, we see that this difference fades away and, by $z\sim5$, the two populations are indistinguishable from one another, in what their median extent is concerned. These results are in agreement with previous findings where both populations are compared \citep[e.g.][]{malhotra2012,jiang2013}.

Our results are consistent with a scenario where a Ly$\alpha$ emitter is a phase through which galaxies may go through in the early stages of their life. From the size evolution perspective, this means that at some point in a galaxy life, when the star-formation is confined to $\lesssim1$ kpc, there are conditions to boost the escape of Ly$\alpha$ photons to our line-of-sight so that we observe the galaxy as a LAE. As time progresses, each galaxy grows in size (along with stellar mass, dust content, metallicity and star formation evolution) and we tend to observe less and less Ly$\alpha$ in emission and observe large galaxies, which are still actively forming stars but do not contribute to the global budget of the observed Ly$\alpha$ emission of the Universe. This decoupling of the two populations with respect to their median size occurs roughly $\sim$1 Gyr after the first galaxies are born. By arguing that the distinction between the two populations happens at the time where a galaxy has evolved for long enough not to be observed as a LAE any more, we can hypothesise that the life cycle of the Ly$\alpha$ emission of a galaxy is typically of the same scale. This means that we may expect on average every galaxy to be observed in Ly$\alpha$ emission for the first $\sim$1 Gyr of its life.

We are aware that our scenario is grossly simplistic, but it finds support by other studies where LAEs are found to be of low mass and low dust content \citep[e.g.][\!\!, see also \citealt{erb2006,kornei2010,hathi2016}]{gawiser2007,pentericci2007,lai2008}. However, there are a number of other studies that report conflicting evidence \citep[e.g.][\!\!\!, see also \citealt{reddy2008}]{finkelstein2009,nilsson2011}. The large diversity of results indicate a more intricate nature of LAEs, pointing to a scenario with possibly recurrent phases of Ly$\alpha$ emission throughout a galaxy's life cycle.

We nonetheless reinforce our findings that LAEs are clearly the most compact population of the two which is consistent with their naturally higher escape fraction of Ly$\alpha$ with respect to an average SFG. At the highest redshifts, the conditions in the Universe were markedly distinct, with most galaxies being very small ($r\lesssim1$ kpc) which in turn renders them more likely to be observed as a LAE, as a consequence of Ly$\alpha$ escaping more easily in smaller galaxies. In the early Universe, typical SFGs have sizes comparable to Ly$\alpha$ emitters, which offers an alternative explanation for the rising fraction of the Ly$\alpha$ emitting SFGs/LBGs up to $z\sim6$ \citep[e.g.][]{hayes2011,stark2011,mallery2012,cassata2015}.

	\subsection{Visual morphology of LAEs}

We show in Figure \ref{fig:LAEs_visual_line} that bright LAEs and high line equivalent width LAEs are more likely to be found with a compact shape. By relating the visual morphology with the structural parameters that we have computed (see Section \ref{section:methodology}) we can find some corroborating signs. Galaxies at the bright end of our LAE sample are found to be smaller, with higher S\'ersic indices, rounder (higher axis-ratio) and with higher light concentrations. These characteristics are relatable to a classical small and round elliptical galaxy which would be classified as compact given our classification scheme. Apart from the discrepancy on the S\'ersic index, we see the same aforementioned trends in the relation of structural parameters with line equivalent width.

	\subsection{The geometric nature of Ly-$\alpha$ emission}\label{ssection:morphlya}

We found some evidence to support that there are some geometric requirements for the successful escape of Ly$\alpha$ photons. In summary, compact and rounded objects are more likely to harbour conditions for such occurrence. This does not invalidate that there are other processes which contribute significantly to such event. The hinted correlations that we find are far from being a definite conclusion on this matter and certainly the existence of outflows, asymmetric or lack of gas and dust distributions can contribute as well to the observation of Ly$\alpha$ in emission. 

This assertion is supported by the relations that we find between light concentration and galaxy axial ratio with Ly$\alpha$ line equivalent width. As found by \citet{sobral2017} the escape fraction of Ly$\alpha$ photons correlates with the Ly$\alpha$ rest-frame equivalent width. Our results are pointing to the fact that galaxies that are rounder and with higher concentrations of their light profile have potentially higher escape fractions. These results are in line with predictions from simulations \citep[e.g][]{verhamme2012}, if we assume the axis-ratio is a good proxy for galaxy inclination with respect to our line of sight. However, a scenario of compact objects with higher volume density of stars can reproduce similar results without the need to invoke galaxy inclination to explain the observations. We believe the latter scenario is more likely the explanation for our results, since we do not expect the majority of our galaxies to have had the time to converge in a rotation supported disk for which the inclination would play a more prevalent role in the perceived escape fraction from our line of sight.


\section{Conclusions}\label{section:conclusions}
			
We present the morphological characterization of a large sample of $\sim$4000 LAEs and quantify their evolution in the first $\sim$3 Gyr of the Universe ($2\lesssim z \lesssim 6$). We study the correlation between the rest-frame UV morphology and the strength of the Ly$\alpha$ emission as a probe to the understanding of the mechanisms underlying the escape of Ly$\alpha$ photons from its host galaxy. Our results can be summarized as:

\begin{itemize}
\item UV sizes of LAEs are constant from $z\sim2$ to $z\sim6$ with sizes of $r_e\sim1.0\pm0.1$ kpc. We observe a rise in sizes towards lower redshifts ($z\sim2$), but the trend is shallow. The little to no evolution seems to hold even down to $z\sim0$.
\item At redshifts $z\lesssim5$, LAEs have sizes that are consistently smaller than those reported for normal SFGs. The difference between the two populations gets more pronounced as we move towards lower redshifts, going from a factor of $\sim$1 at $z\gtrsim5$ to SFGs being a factor of $\sim2-4$ larger than LAEs for $z\lesssim2$. We hypothesize that the small/compact nature of LAEs is potentially linked to physical escape mechanisms of Ly$\alpha$ photons. In the early Universe, typical SFGs have sizes comparable to Ly$\alpha$ emitters which offers an alternative explanation for the rising fraction of the Ly$\alpha$ emitting SFGs/LBGs up to $z\sim6$.
\item The profiles of LAEs as seen from the rest-frame UV are remarkably constant from $z\sim2$ up to $z\sim6$ with $n\sim1.7$ being slightly steeper than a pure exponential disk. The same scenario is seen in the evolution of the light concentration and axis ratio of LAEs. 
\item When looking at subsamples of fixed Ly$\alpha$ luminosity and fixed Ly$\alpha$ equivalent width we find a more pronounced evolution with redshift for S\'ersic indexes and light concentration (but not for galaxy sizes). LAEs with the lowest EWs (which are the brightest in UV within our sample) are those who show a stronger evolution. This is likely happening because low EW LAEs are those more similar to regular LBGs/SFGs \citep[e.g.][]{hathi2016}, which show some evolution of their light profiles \citep[e.g.][]{morishita2014}.
\item We find that most LAEs in our sample are compact in their morphology. The fraction of compact LAEs is larger at high line equivalent widths and also at high Ly$\alpha$ luminosity.
\item Ly$\alpha$ equivalent width seems to correlate stronger with the axis ratio and size of galaxies than any other morphological parameter we have tested. Strong LAEs are found more likely in small and rounder galaxies ($r_e\sim0.8$ kpc and $b/a\sim0.5$).
\item The results  that we report as the median properties of the population are corroborated by the morphological properties of the stacked profiles of LAEs. This means that even when the image depth is increased, we find no difference with respect to the detected LAEs and discard the existence of an extended lower surface brightness region around UV-bright LAEs.
\end{itemize} 

In broad terms, our results provide a global picture on the rest-frame UV morphology of LAEs in the early Universe. We find that this particular population of galaxies does not evolve significantly in the first 3 Gyr of the Universe and that it departs from the evolution of normal star-forming galaxies for $z<4$, in what galaxy sizes is concerned.


\section*{Acknowledgements}

We are grateful to the referee's useful and constructive comments and suggestions which improve this paper.
This work was supported by Funda\c{c}\~{a}o para a Ci\^{e}ncia e a Tecnologia (FCT) through the research grant UID/FIS/04434/2013. ASP, PhD::SPACE fellow, acknowledges support from the FCT through the fellowship PD/BD/52706/2014. DS acknowledges financial support from the Netherlands Organisation for Scientific research (NWO) through a Veni fellowship and from Lancaster University through an Early Career Internal Grant A100679.
The authors acknowledge the award of time from programmes: I13AN002, I14AN002, 088-INT7/14A, I14BN006, 118-INT13/14B, I15AN008 on the Isaac Newton Telescope (INT). INT is operated on the island of La Palma by the Isaac Newton Group in the Spanish Observatorio del Roque de los Muchachos of the Instituto de Astrofisica de Canarias.
We are grateful to the CFHTLS, COSMOS-UltraVISTA, UKIDSS, SXDF and COSMOS survey teams. Without these legacy surveys, this research would have been impossible.
We thank the VUDS team for making available spectroscopic redshifts from data obtained with VIMOS at the European Southern Observatory Very Large Telescope, Paranal, Chile, under Large Program 185.A-0791.
This work was only possible by the use of the following \textsc{python} packages: NumPy \& SciPy \citep{walt2011,jones2001}, Matplotlib \citep{hunter2007} and Astropy \citep{robitaille2013}.


\bibliographystyle{mnras}
\bibliography{refs}


\appendix

\section{Ly$\alpha$ emitters stacks}\label{section:appendixA}

\begin{figure}
\centering
\includegraphics[width=\linewidth]{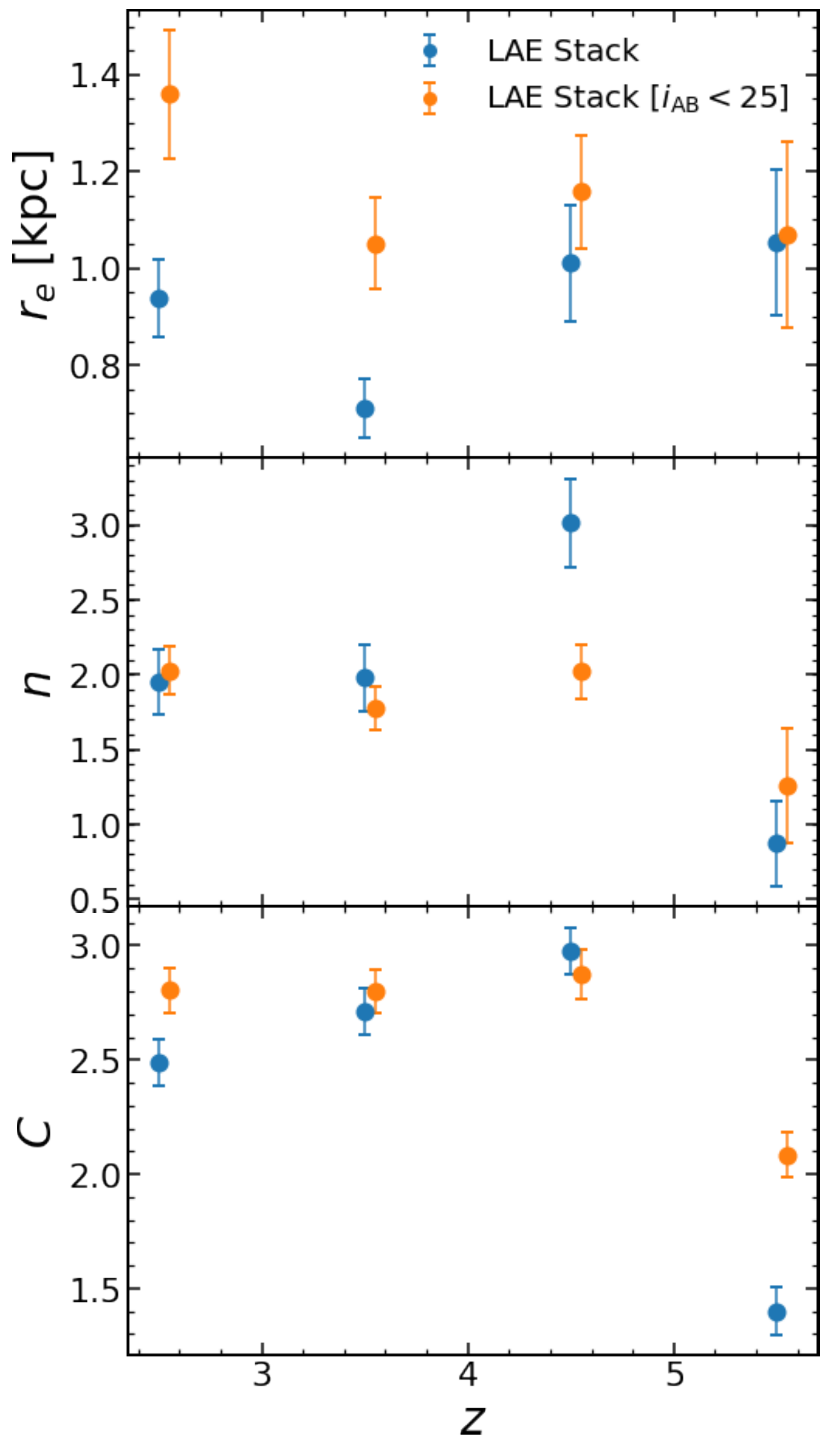}
\caption[LAE values for different stack samples at $2\lesssim z\lesssim 6$]{LAE values for different stack samples at $2\lesssim z\lesssim 6$. From top to bottom we show the derived values for the stack of the full sample (in blue) and the stack of the $i_\mathrm{AB}<25$ sample (in orange).}
\label{fig:LAEs_stack_corrections}
\end{figure}

\begin{figure*}
\centering
\includegraphics[width=\linewidth]{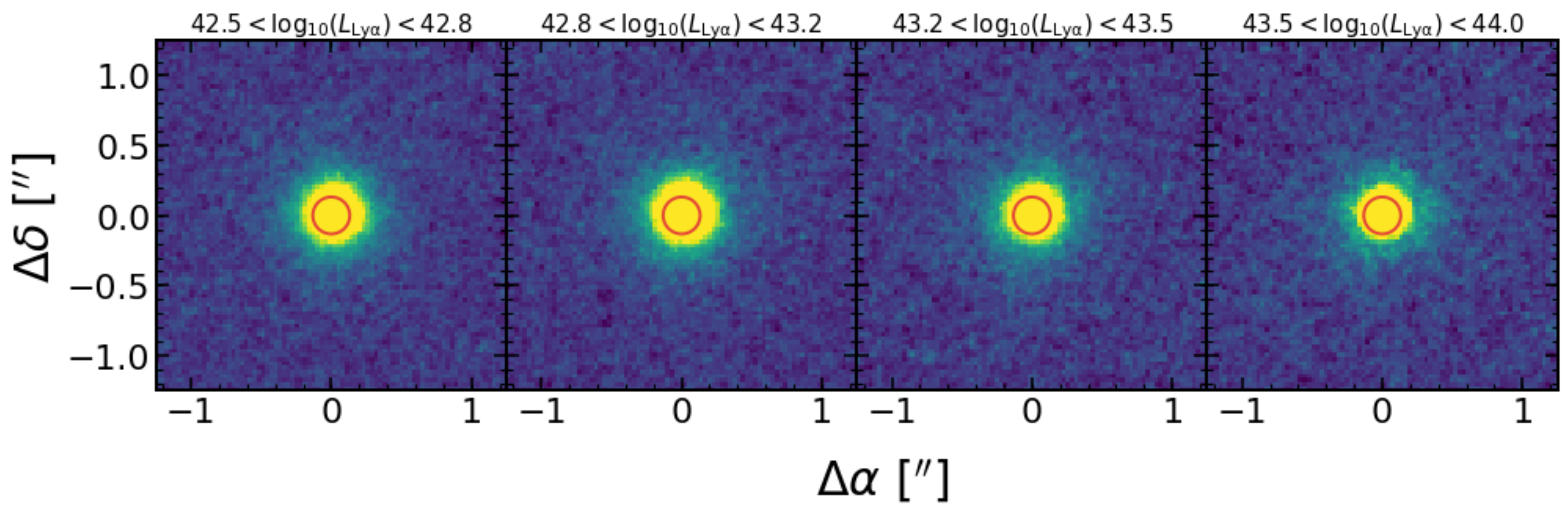}
\includegraphics[width=\linewidth]{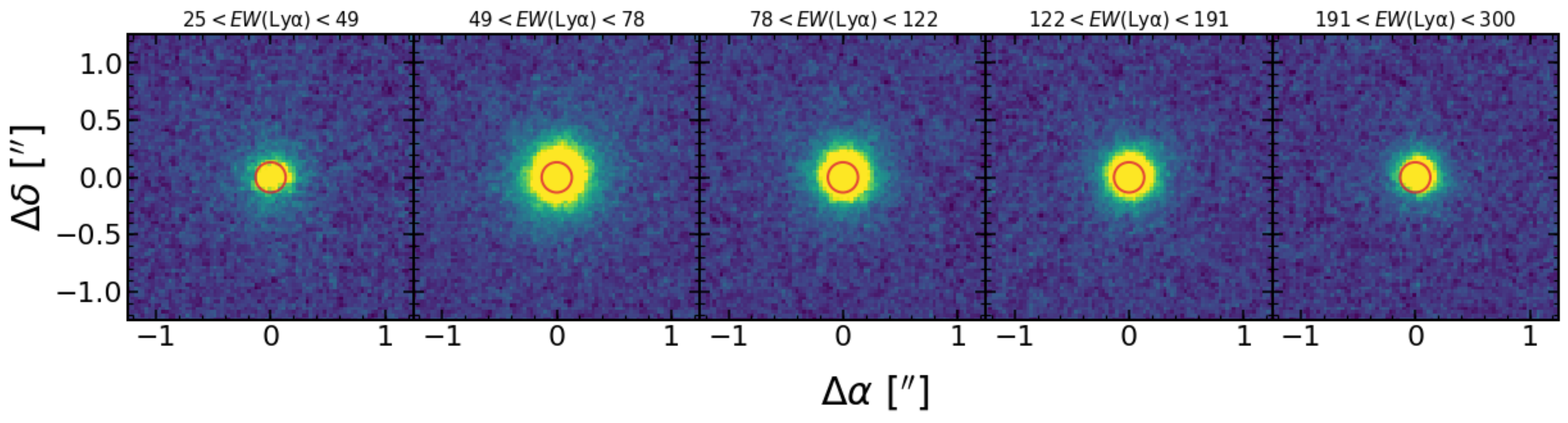}
\caption[LAE image stacks at $2\lesssim z\lesssim 6$]{Examples of LAE stacks for each of the bins that we use in this study in terms of Ly$\alpha$ luminosity (top) and Ly$\alpha$ equivalent width (bottom). In each panel, the intensity levels range from -3$\sigma_\mathrm{sky}$ to 15$\sigma_\mathrm{sky}$, where $\sigma_\mathrm{sky}$ is the sky rms. The red circle in each panel has a physical radius of 1 kpc.}
\label{fig:LAEs_stack_examplesAA}
\end{figure*}

To correct for the possible biases on morphological parameters induced by combining astrometric errors and Ly$\alpha$-UV mismatch (see Section \ref{section:method_stacks}) we have computed image stacks using only a subset of galaxies with $i_\mathrm{AB}<25$ and compare that to the full sample. Then we compute the corrections to be applied to the measured stack values as the ratio between these two quantities. We show in Figure \ref{fig:LAEs_stack_corrections} the values we get for three different morphological quantifiers in the case of the full sample and the $i_\mathrm{AB}<25$ sample.

\section{General properties of individual samples}\label{sec:general_properties}

\begin{figure*}
\centering
\includegraphics[width=\linewidth]{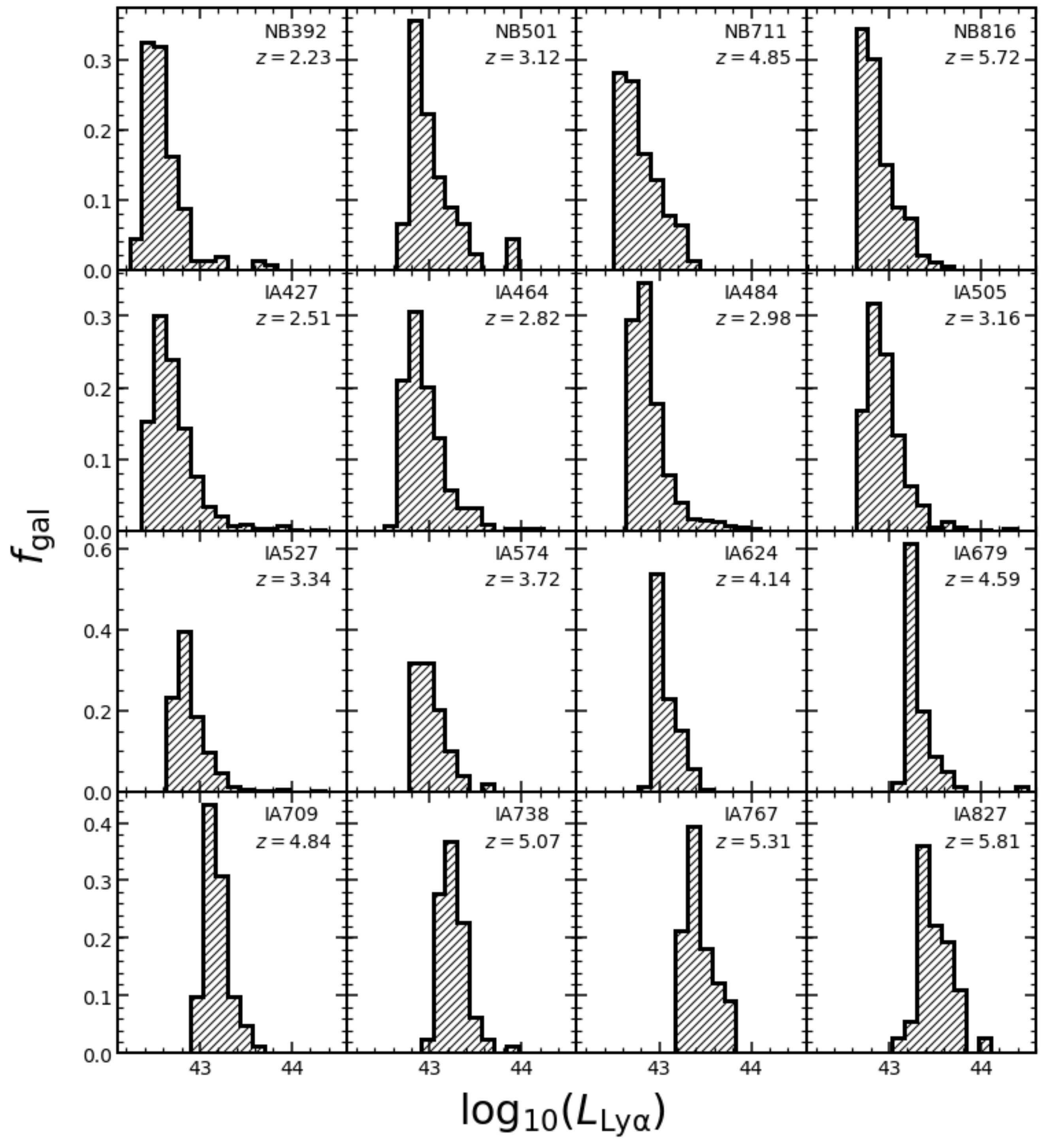}
\caption[LAE luminosities for all NB and IB samples]{Distribution of Ly$\alpha$ luminosities for all LAE candidates in our samples. NB samples are shown in the top row and IB samples in the other three rows. We find higher Ly$\alpha$ luminosities with increasing redshift and that NB samples probe fainter LAEs when compared to IB samples at similar redshifts.}
\label{fig:LAEs_samples_lum}
\end{figure*}

\begin{figure*}
\centering
\includegraphics[width=\linewidth]{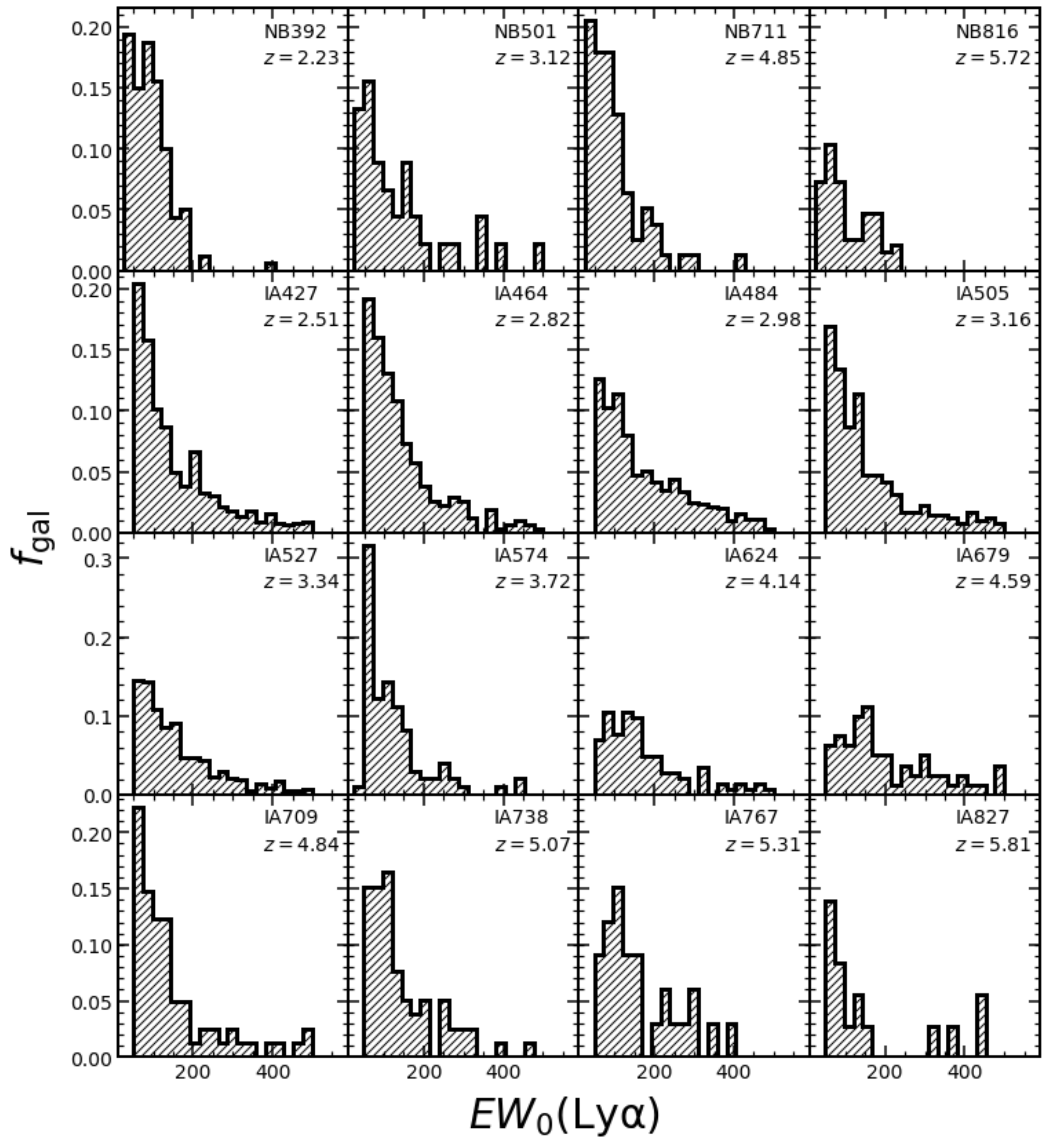}
\caption[LAE Ly$\alpha$ equivalent widths for all NB and IB samples]{Distribution of Ly$\alpha$ equivalent widths for all LAE candidates in our samples. NB samples are shown in the top row and IB samples in the other three rows. Despite the different line equivalent width cuts for NB and IB samples, we find similar distributions for all bands independent of the redshift.}
\label{fig:LAEs_samples_ew}
\end{figure*}

\begin{figure*}
\centering
\includegraphics[width=\linewidth]{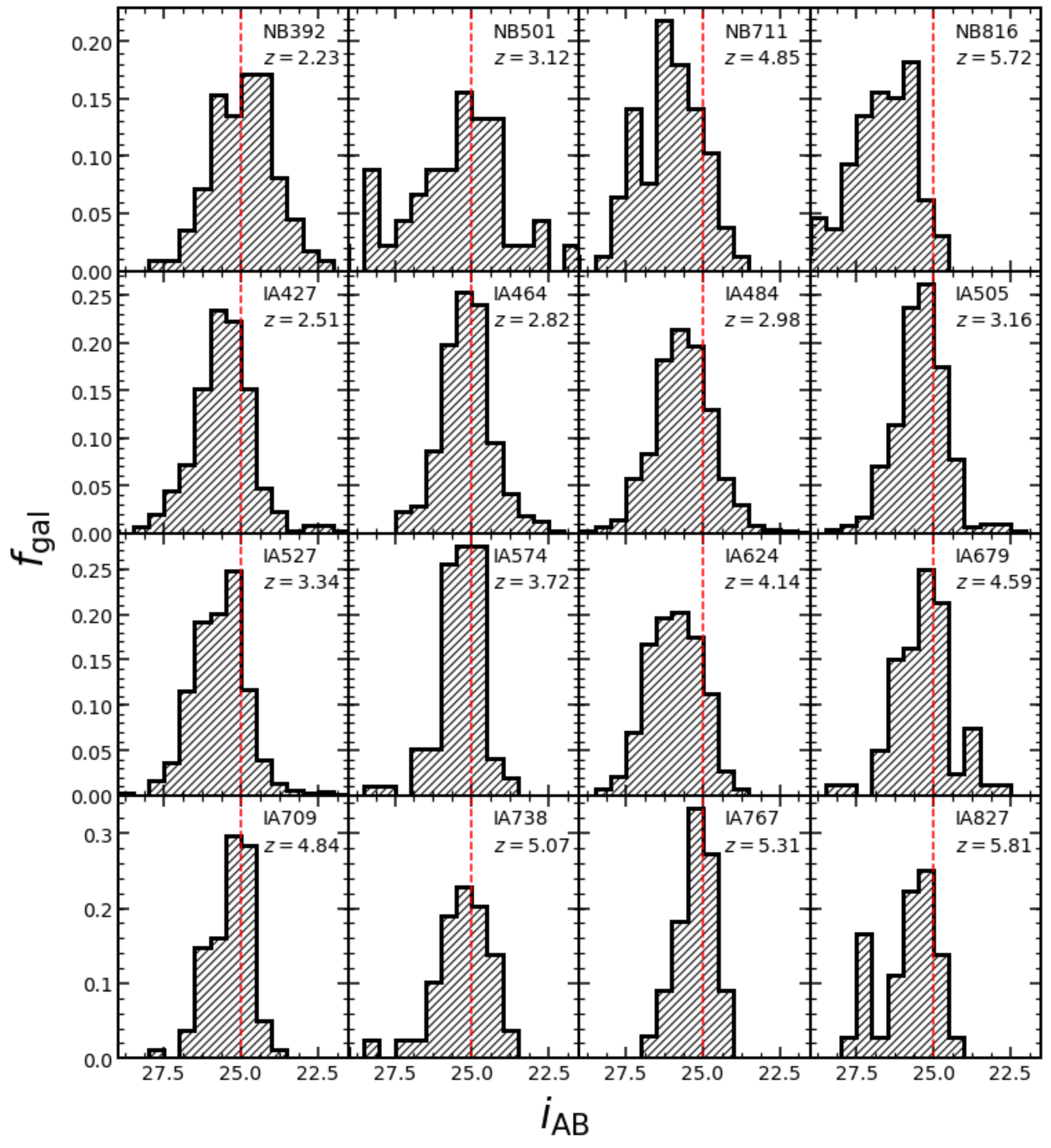}
\caption[LAE $i_\mathrm{AB}$ magnitudes for all NB and IB samples]{Distribution of $i_\mathrm{AB}$ for all LAE candidates in our samples. NB samples are shown in the top row and IB samples in the other three rows. The vertical red dashed line highlights the $i_\mathrm{AB}=25$ limit that we use in this paper. There is a large fraction of LAEs fainter than our imposed limit and this is more severe when considering the higher redshift samples.}
\label{fig:LAEs_samples_mag}
\end{figure*}

We show in Figures \ref{fig:LAEs_samples_lum}-\ref{fig:LAEs_samples_mag} the distribution of Ly$\alpha$ luminosities, Ly$\alpha$ equivalent widths and $i_\mathrm{AB}$ continuum magnitudes for each individual sample present in this paper. In terms of Ly$\alpha$ luminosity, we see a clear dependence of the minimum detected luminosity with redshift as expected by a constant line flux limit that is imposed by the observations. In terms of Ly$\alpha$ equivalent width all bands probe nearly the same region. We note however that for NB samples we are able to go down to $EW>25$\ \AA\ while for IB samples we have the lower limit at $EW>50$\ \AA. Finally, for $i_\mathrm{AB}$ we find a similar trend as for the Ly$\alpha$ luminosity with fainter galaxies observed at higher redshifts.

\section{LAE stacks for a multi-parameter exploration}

\begin{figure*}
\centering
\includegraphics[width=\linewidth]{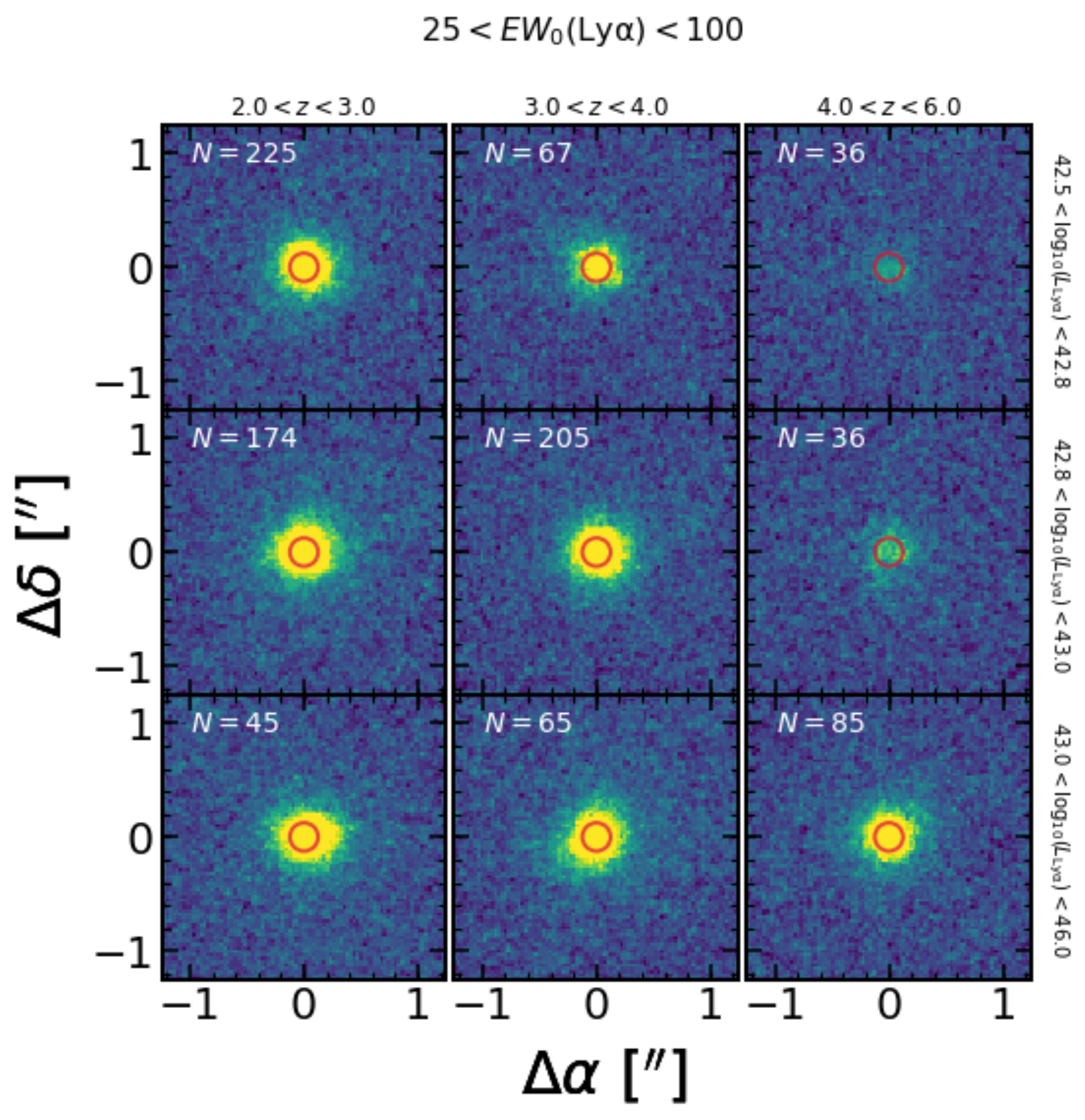}
\caption[Individual stacks for multi-parameter exploration]{Individual stacks for multi-parameter exploration with $25<EW_0(\mathrm{Ly\alpha})<100$\ \AA. Each column represents a different redshift bin and each row a different luminosity bin. The red circle has a 1 kpc radius at the median redshift of each bin.}
\label{fig:LAEs_cubeStack1}
\end{figure*}

\begin{figure*}
\centering
\includegraphics[width=\linewidth]{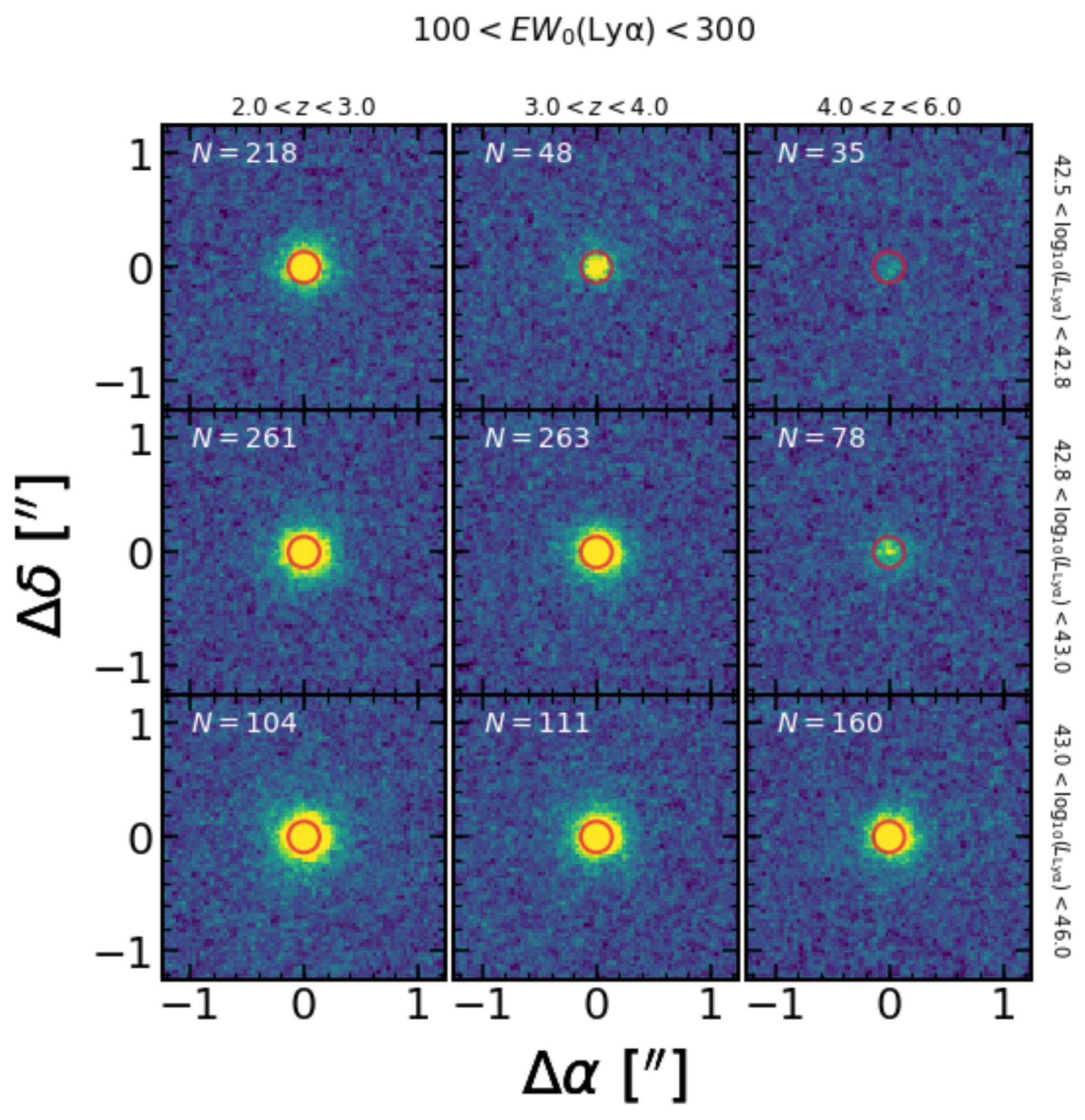}
\caption[Individual stacks for multi-parameter exploration]{Individual stacks for multi-parameter exploration with $100<EW_0(\mathrm{Ly\alpha})<300$\ \AA. Each column represents a different redshift bin and each row a different luminosity bin. The red circle has a 1 kpc radius at the median redshift of each bin.}
\label{fig:LAEs_cubeStack2}
\end{figure*}

\begin{figure*}
\centering
\includegraphics[width=\linewidth]{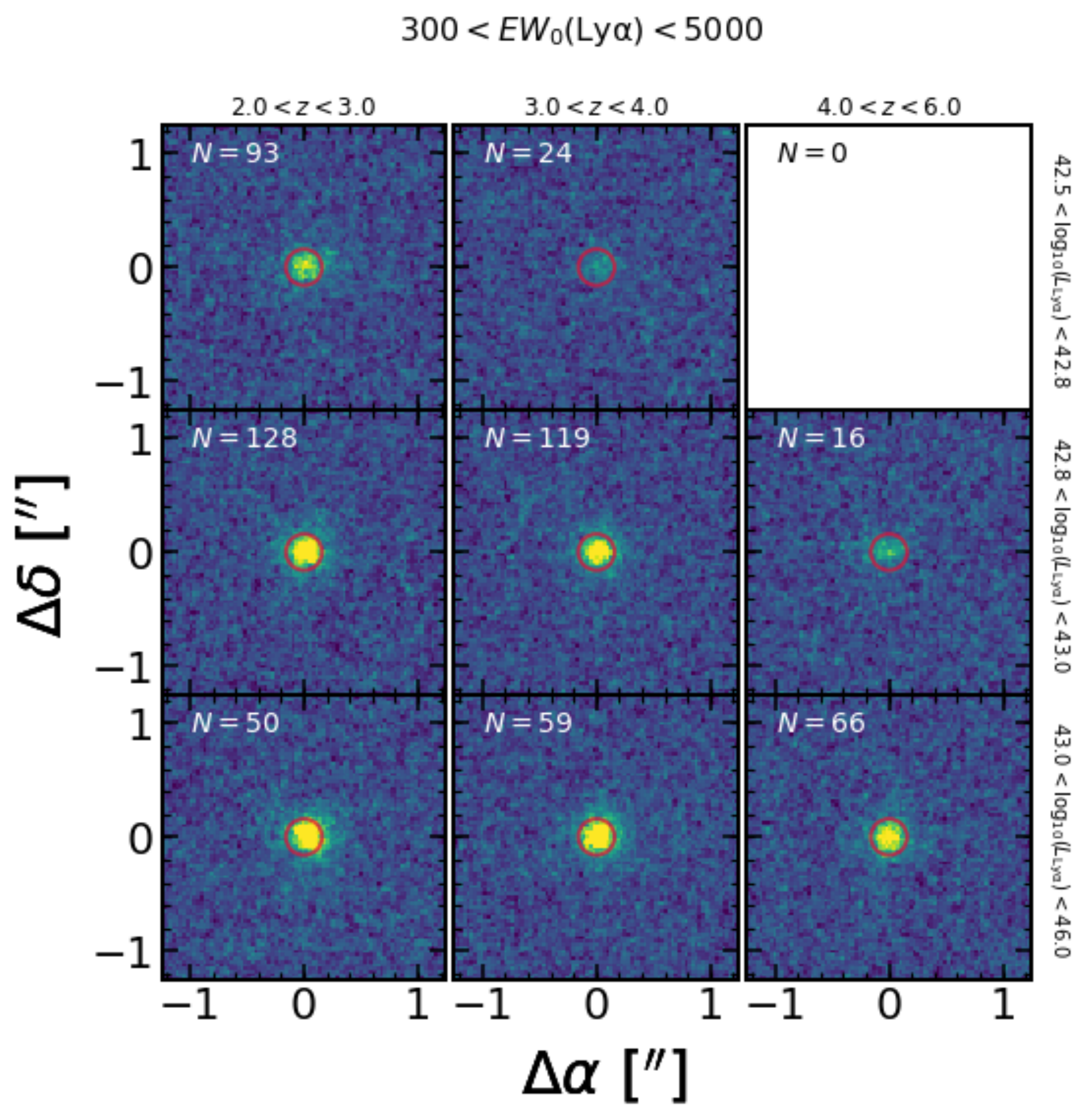}
\caption[Individual stacks for multi-parameter exploration]{Individual stacks for multi-parameter exploration with $300<EW_0(\mathrm{Ly\alpha})<5000$\ \AA. Each column represents a different redshift bin and each row a different luminosity bin. The red circle has a 1 kpc radius at the median redshift of each bin. The empty slot in the upper right corner is due to a lack of galaxies in our sample occupying that bin.}
\label{fig:LAEs_cubeStack3}
\end{figure*}

This section contains images showing the resulting stacks for studying independently the redshift, Ly$\alpha$ luminosity and Ly$\alpha$ equivalent width dependence of rest-frame UV LAE morphology (Figures \ref{fig:LAEs_cubeStack1}-\ref{fig:LAEs_cubeStack3}).

\bsp
\label{lastpage}
\end{document}